\documentclass[twocolumn,aps,superscriptaddress,prc,nofootinbib,showpacs,showkeys]{revtex4-2}
\usepackage{graphicx}
\usepackage{amssymb}
\usepackage{amsmath}
\usepackage{amsthm}
\usepackage{color}
\usepackage[normalem]{ulem}
\usepackage{slashed}
\usepackage{mathrsfs}
\usepackage{hyperref}
\hypersetup{
	colorlinks=true,        
	linkcolor=blue,         
	citecolor=cyan,         
}

\newcommand{\fig}{Fig.~}

\allowdisplaybreaks



\begin{document}
	
\title{Evolution of global polarization in relativistic heavy-ion collisions within a perturbative approach}
\author{Xiaowen Li}
\affiliation{Institute of Frontier and Interdisciplinary Science, Shandong University, Qingdao, Shandong, 266237, China}

\author{Ze-Fang Jiang}
\affiliation{Department of Physics and Electronic-Information Engineering, Hubei Engineering University, Xiaogan, Hubei, 432000, China}
\affiliation{Institute of Particle Physics and Key Laboratory of Quark and Lepton Physics (MOE), Central China Normal University, Wuhan, Hubei, 430079, China}

\author{Shanshan Cao}
\email{shanshan.cao@sdu.edu.cn}
\affiliation{Institute of Frontier and Interdisciplinary Science, Shandong University, Qingdao, Shandong, 266237, China}

\author{Jian Deng}
\email{jdeng@sdu.edu.cn}
\affiliation{Institute of Frontier and Interdisciplinary Science, Shandong University, Qingdao, Shandong, 266237, China}

	
\begin{abstract}
Extremely large angular orbital momentum can be produced in non-central heavy-ion collisions, leading to a strong transverse polarization of partons that scatter through the quark-gluon plasma (QGP) due to spin-orbital coupling. We develop a perturbative approach to describe the formation and spacetime evolution of quark polarization inside the QGP. Polarization from both the initial hard scatterings and interactions with the QGP have been consistently described using the quark-potential scattering approach, which has been coupled to realistic initial condition calculation and the subsequent (3+1)-dimensional viscous hydrodynamic simulation of the QGP for the first time. Within this improved approach, we have found that different spacetime-rapidity-dependent initial energy density distributions generate different time evolution profiles of the longitudinal flow velocity gradient of the QGP, which further lead to an approximately 15\% difference in the final polarization of quarks collected on the hadronization hypersurface of the QGP. Therefore, in addition to the collective flow coefficients, the hyperon polarization may serve as a novel tool to help constrain the initial condition of the hot nuclear matter created in high-energy nuclear collisions.
\end{abstract}	
\maketitle	
	
\section{introduction}	
\label{sec1}

High-energy nucleus-nuclues collisions at the Relativistic Heavy-Ion Collider (RHIC) and the Large Hadron Collider (LHC) create a color deconfined state of nuclear matter, known as the Quark-Gluon Plasma (QGP), whose strongly-coupled nature has been confirmed by both the collective flow of soft hadrons and the quenching phenomena of jets observed in these energetic collisions~\cite{Gyulassy:2004zy,Jacobs:2004qv}. In addition to the hottest and densest environment one may obtain in laboratory, non-central heavy-ion collisions also deposit huge amount of angular momentum (on the order of $10^5$) into the nuclear matter~\cite{Gao:2007bc}, leading to the most vortical system (vorticity on the order of $10^{21}$~s$^{-1}$) one may create~\cite{Deng:2016gyh,Jiang:2016woz}.


It was proposed in Ref.~\cite{Liang:2004ph} that this large angular momentum can generate polarization of quarks via their spin-orbital interactions, and in the end be observed as the global polarization of hyperons, e.g. $\Lambda$, $\Sigma$ and $\Xi$. Other possible observables include vector meson spin alignment~\cite{Liang:2004xn} and the emission of circularly polarized photons~\cite{Ipp:2007ng}. On the experimental side, search for the $\Lambda$ polarization in heavy-ion collisions was initiated by the STAR Collaboration~\cite{STAR:2007ccu} and recently confirmed in Refs.~\cite{STAR:2017ckg,STAR:2018gyt}. The non-trivial dependences of the $\Lambda$ polarization on the collision energy, rapidity region and the transverse momentum of $\Lambda$ have attracted tremendous theoretical efforts on exploring the detailed mechanisms that generate polarization~\cite{Becattini:2020ngo}. 

A major category of theoretical approaches is based on the assumption that the spin degrees of freedom are at local equilibrium on the hadronization hypersurface of the QGP~\cite{Becattini:2007nd,Becattini:2007sr}, allowing one to extract the polarization of the final-state hadrons from the extended Cooper-Frye formalism for particles with spin. The key quantity that drives the development of polarization is the thermal vorticity~\cite{Becattini:2007nd,Becattini:2013fla}. The evolution of the QGP can be either simulated with a perturbative-based transport model (e.g. APMT)~\cite{Li:2017slc,Li:2021zwq,Huang:2020dtn} or be considered as a strongly-coupled medium (e.g. hydrodynamics)~\cite{Betz:2007kg,Pang:2016igs,Fang:2016vpj,Karpenko:2016jyx,Baznat:2017jfj,Xie:2017upb,Florkowski:2017ruc,Hattori:2019lfp,Xie:2019wxz,Fukushima:2020ucl,Li:2020eon,Fu:2020oxj,Bhadury:2021oat,Ryu:2021lnx,Wu:2022mkr}. While many of these studies provide a reasonable description of the experimental data, in order to address non-equilibrium effects on polarization, it is necessary to introduce the quantum kinetic theory. This have been recently developed in Refs.~\cite{Sun:2017xhx,Liu:2019krs,Wang:2020pej,Yang:2020hri,Weickgenannt:2020aaf}, where crucial questions like conservation of angular momentum in a transport model and evolution towards local equilibrium have been explored. Instead of a full transport description of interactions between quark pairs, an alternative microscopic approach was proposed in Ref.~\cite{Huang:2011ru}, where each quark is considered interacting with a mean field of the medium background. The perturbative scattering picture from the earliest study~\cite{Liang:2004ph} has been extended to multiple scatterings of quarks through the QGP, from which one can observe the development of the quark polarization driven by the longitudinal velocity gradient of the fluid background.

In the present study, we will further develop this perturbative approach~\cite{Huang:2011ru} in several aspects. We will consistently apply this perturbative scattering picture to both the initial hard collisions between nuclei and quark scatterings through the QGP for the first time, which allows us to study the initial production of polarization and its subsequent evolution within the same framework. A simplification was usually applied in earlier perturbative calculations~\cite{Liang:2004ph,Huang:2011ru} where a projectile quark is constrained in a half hemisphere relative to the target potential. Realistic spatial distributions of projectile and target will be taken into account for a more precise evaluation of the average quark polarization in this work. In addition, a (3+1)-D viscous hydrodynamic model CLVisc~\cite{Pang:2018zzo,Wu:2018cpc} will be introduced to provide a realistic evolution of the longitudinal flow velocity profile of the QGP. Impacts of using different initial conditions for the hydrodynamic expansion on the final-state global polarization will be investigated in detail. We will focus on the global polarization that results from the large angular momentum with a direction perpendicular to the reaction plane of non-central heavy-ion collisions. The other crucial branch of polarization, the local (or longitudinal) polarization due to the QGP expansion in the transverse plane~\cite{STAR:2019erd,Fu:2021pok,Becattini:2021iol} is beyond the scope of this work.

The rest of this paper will be organized as follows. In Sec.~\ref{sec2}, we will discuss the quark polarization produced by the initial hard collisions, and investigate how it is affected by the nucleon density distribution. In Sec.~\ref{sec3}, we will study the evolution of the global polarization through the QGP phase, and explore how the final-state polarization depends on the initial energy density profile of the hydrodynamic evolution of the QGP. A summary will be presented in Sec.~\ref{sec4}.

\section{Initial polarization from hard scatterings}
\label{sec2}

\begin{figure}[tbp]
	\centering
	\includegraphics[width=0.4\textwidth]{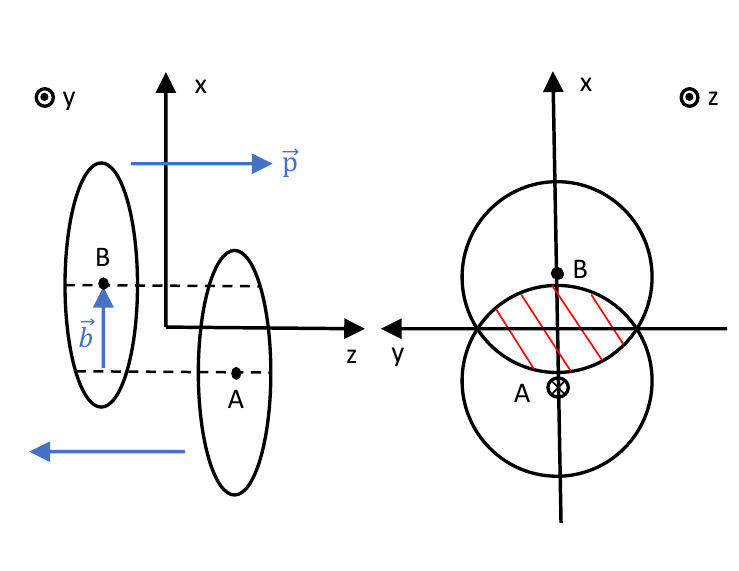}
	\addtocounter{figure}{0}
	\renewcommand{\figurename}{FIG}
	\caption{(Color online) Side view (left panel) and top view (right panel) on a non-central heavy-ion collision event.}
	\label{qgp}
\end{figure}

We consider a non-central collision event between two nuclei, as demonstrated in Fig.~\ref{qgp}. The momentum ($\vec{p}$) of each nucleon from the projectile nucleus (B) is assigned along the $+\hat{z}$ direction, while the momentum of each nucleon from the target (A) is the opposite. The impact parameter ($\vec{b}$), pointing from the target to the projectile, is assigned along the $+\hat{x}$ direction. The reaction plane is then defined as the $z$-$x$ plane, whose direction can be determined as $\hat{n} = +\hat{y}$. This non-central collision would deposit huge angular momentum into the overlapping region between the two nuclei, leading to the polarization of quarks that constitute the QGP medium via the spin-orbital coupling. In this section, we study the formation of the polarization from this initial hard nucleus-nucleus scattering, and discuss how the magnitude of polarization depends on the initial nucleus geometry.

\begin{figure}[tbp]
	\centering
	\includegraphics[width=0.28\textwidth]{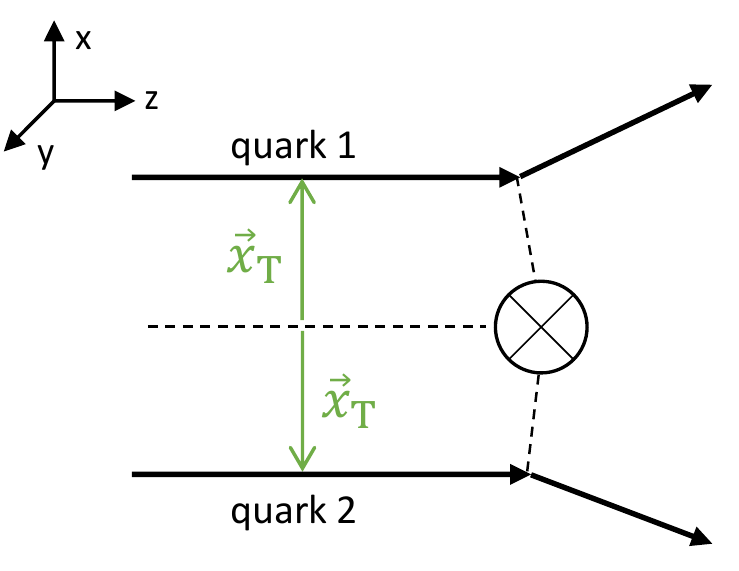}
	\caption{(Color online) Cartoon for the static potential scattering process.}
	\label{model}
\end{figure}

Following Ref.~\cite{Liang:2004ph}, we investigate a scattering between a quark and a static potential $A^0(q_\mathrm{T})=g/(q_\mathrm{T}^2+\mu^2)$, where $g^2=\alpha_\mathrm{s}/(4\pi)$ is the strong coupling constant, $q_\mathrm{T}$ represents the momentum transfer and $\mu$ is Debye screening mass of the exchanged gluon. This potential scattering picture can be consistently applied to the initial hard scattering and the subsequent quark scattering through the QGP medium. For the initial hard scattering in vacuum, we take $\mu=0.5$~GeV as in Ref.~\cite{Liang:2004ph}; while for the in-medium scattering, $\mu^2=g^2(N_c+N_f/2)T^2/3$ will be adopted in the next section, with $T$ being the local temperature of the QGP and $N_c=N_f=3$ in the present study. For a given initial momentum $(E,\vec{p})$ and a final spin $\lambda/2$ of the outgoing quark along $\hat{n}$, the differential cross section for an initially unpolarized quark can be obtained as~\cite{Liang:2004ph}
\begin{equation}
\label{eq:xsection0}
\frac{d^2 \sigma_{\lambda}}{d^2x_\mathrm{T}}= C_\mathrm{T} \int \frac{d^2q_\mathrm{T}}{(2\pi)^2} \frac{d^2k_\mathrm{T}}{(2\pi)^2} e^{ i( \vec{k}_\mathrm{T}- \vec{q}_\mathrm{T} ) \cdot \vec{x}_\mathrm{T}  } \mathcal{I}_\lambda,
\end{equation}
with
\begin{align}
\label{eq:I0}
\mathcal{I}_\lambda&= \frac{1}{2}\sum_{\lambda_i}\mathcal{I}_{\lambda\lambda_i}\nonumber\\ 
&=\frac{g^2}{2(2E)^2}\sum_{\lambda_i} \bar{u}_{\lambda} ( p_q ) \slashed{A}(q_\mathrm{T}) u_{\lambda_i}(p) \bar{u}_{\lambda_i}(p) \slashed{A}( k_\mathrm{T} )	u_{\lambda}( p_k )\nonumber\\
&=\frac{g^2}{2(2E)^2}\bar{u}_{\lambda} ( p_q ) \slashed{A}(q_\mathrm{T}) (\slashed{p}+m) \slashed{A}( k_\mathrm{T} )	u_{\lambda}( p_k ).
\end{align} 
In the above equations, $C_\mathrm{T}=2/9$ is the color factor associated with the target, $A^\mu=(A^0,\vec{0})$ is the scattering potential, $\vec{q}_\mathrm{T}$ ($\vec{k}_\mathrm{T}$) is the momentum transfer between the quark and the potential in the real (complex conjugate) space, $\vec{x}_\mathrm{T}$ is the relative transverse distance pointing from the scattering center to the quark (as illustrated in Fig.~\ref{model}), and $m$ is the quark mass. After being scattered, the outgoing quark momentum is $\vec{p}_{q(k)}=\vec{p}+\vec{q}_\mathrm{T}(\vec{k}_\mathrm{T})$. Here, we have averaged over the initial spin states ($\lambda_i$) for the unpolarized projectile quark.

For a high energy quark, one may assume a small scattering angle ($q_\mathrm{T},k_\mathrm{T}\sim \mu\ll E$), which simplifies the $\mathcal{I}_\lambda$ part as
\begin{equation}
\label{eq:I0simplify}
\frac{\mathcal{I}_\lambda}{g^2}\approx \frac{1}{2}A_0(q_\mathrm{T})A_0(k_\mathrm{T})\left[1-i\lambda\frac{(\vec{q}_\mathrm{T}-\vec{k}_\mathrm{T})\cdot (\hat{n}\times\vec{p})}{2E(E+m)}\right].
\end{equation}
The first part in the square bracket is spin-independent, which contributes to the total cross section as 
\begin{equation}
\label{eq:1xsectionTotal0}
\frac{d \sigma}{d^2x_\mathrm{T}}  = \frac{d \sigma_+}{d^2x_\mathrm{T}} + \frac{d \sigma_-}{d^2x_\mathrm{T}} = 4C_\mathrm{T}\alpha^2_\mathrm{s}K^2_0(\mu x_\mathrm{T});
\end{equation}
while the second term is spin-dependent, contributing to the difference between positive and negative spin configurations as 
\begin{align}
\label{eq:1xsectionDiff0}
\frac{d \Delta \sigma}{d^2x_\mathrm{T}}   &= \frac{d \sigma_+}{d^2x_\mathrm{T}} - \frac{d \sigma_-}{d^2x_\mathrm{T}}\\ \nonumber
	&= -4C_\mathrm{T}\alpha^2_\mathrm{s} \mu \frac{ \vec{p} \cdot (\hat{x}_\mathrm{T} \times \hat{n} )}{E(E+m)} K_0(\mu x_\mathrm{T})K_1(\mu x_\mathrm{T}).
\end{align}
Here, $\hat{x}_\mathrm{T}=\vec{x}_\mathrm{T}/x_\mathrm{T}$ and $K_n$'s are the modified Bessel functions. The polarization of quarks is then defined as the ratio between the integrated $\Delta \sigma$ and $\sigma$ above, $P=\Delta \sigma/\sigma$. Although this ratio does not explicitly depend on $\alpha_\mathrm{s}$ (or $g$), the strong coupling constant will affect the magnitude of polarization through the Debye screening mass, as will be discussed later in this work. When using the above equation to evaluate the initial polarization, we take the center-of-mass momentum $p=100$~GeV along $\pm\hat{z}$ directions for Au-Au collisions at $\sqrt{s_\mathrm{NN}}=200$~GeV, and the quark mass $m=100$~MeV for estimating their energy $E$. We have verified that varying the quark mass within $5\sim 300$~MeV has no visible impact on the polarization result.

Observed from Eq.~(\ref{eq:1xsectionDiff0}), the sign of the cross section difference depends on the relative position between the projectile quark and the scattering center. As sketched in Fig.~\ref{model}, if the projectile quark (e.g. quark 1) is in the $x>0$ hemisphere with respective to the scattering center, $d \Delta \sigma / d^2x_\mathrm{T}$ is negative; contrarily, for the $x<0$ hemisphere (e.g. quark 2),  $d \Delta \sigma / d^2x_\mathrm{T}$ is positive. We note that in earlier studies~\cite{Liang:2004ph,Huang:2011ru}, a simplified assumption is adopted for two-particle scattering, where one quark from B is always in the $x>0$ hemisphere relative to the other quark from A. By integrating over the corresponding half hemisphere, a heuristic approximation for the quark polarization is obtained as $P=-\pi\mu p/[2E(E+m)]$.\footnote{If one integrates over the entire $x_\mathrm{T}$ plane instead, zero polarization will be obtained.} To improve the estimation of polarization from non-central nuclear collisions, we extend this two-particle scattering picture with relative position restricted in the half hemisphere to scatterings between all pairs of nucleons that are realistically distributed in the two colliding nuclei. We will show that the finite polarization in the end is not because one quark from B is {\it always} above ($x>0$) the other quark from A, but because the nuclear matter contributed by B is {\it on average} above that by A.

To take into account the configurations of nuclei A and B, we need to integrate Eqs.~(\ref{eq:1xsectionTotal0}) and~(\ref{eq:1xsectionDiff0}) over the whole transverse plane, weighted by the thickness functions of A and B. Two different models of the nuclear matter distribution are used and compared here. The first one is the Hard Sphere Model, where the nuclear matter is assumed to be strictly constrained within a radius $R$. The corresponding thickness function reads
\begin{equation}
\label{eq:thicknessHS}
	\begin{aligned}
		{ T_\mathrm{A,B}^\mathrm{HS}(x,y)} &= \frac{3A}{2 \pi R^3} \sqrt{R^2-(x\pm b/2)^2-y^2} \\ 
		& \times { \Theta(R-\sqrt{(x\pm b/2)^2+y^2})},
	\end{aligned}
\end{equation}
where the centers of nuclei A and B are placed at $(-b/2,0)$ and $(b/2,0)$ in the transverse plane respectively, and $A$ on the right hand side denotes the nucleon number (same for A and B in this work). Another model for the nuclear matter density is the Woods-Saxon distribution, written as 
\begin{equation}
\label{eq:thicknessWS}
	{ T_\mathrm{A,B}^\mathrm{WS}(x,y)}= \int dz \frac{\rho_0} {1 + e^{\left(\sqrt{(x\pm b/2)^2+y^2+z^2}-R\right)/a} },
\end{equation}
where $\rho_0$ is the equilibrium density of nuclear matter, and $a$ is the surface thickness parameter. Unlike the Hard Sphere Model, the nuclear matter density in the Woods-Saxon model decreases smoothly, though rapidly, when the position is over $R+a$ away from the nucleus center.

Since only nucleons that participate in the initial nucleus-nucleus collisions contribute to the QGP medium, we only include those participant nucleons in our calculation. For the above two models, the participant number density read~\cite{Gao:2007bc}
\begin{align}
\label{eq:partHS}
&T^\mathrm{HS}_{1,2}(x,y) = T_\mathrm{A,B}^\mathrm{HS}\Theta(R-\sqrt{(x\mp b/2)^2+y^2}),\\
\label{eq:partWS}
&T^\mathrm{WS}_{1,2}(x,y) = T^\mathrm{WS}_\mathrm{A,B}\left[1-\exp (-\sigma_\mathrm{NN} \times T^\mathrm{WS}_\mathrm{B,A})\right],
\end{align}
respectively, in which $\sigma_\mathrm{NN}$ is the inelastic cross section of nucleon-nucleon collision, and the subscript 1(2) is used to represent participants from nucleus A(B). For Au-Au collisions at $\sqrt{s_\mathrm{NN}}=200$~GeV, parameters in Eqs.~(\ref{eq:thicknessHS}) - (\ref{eq:partWS}) are set as $A=197$, $R=6.38$~fm, $\rho_0 = 0.17$~fm$^{-3}$, $a = 0.535$~fm and $\sigma_\mathrm{NN} = 42$~mb~\cite{Loizides:2017ack}.

With these participant number density distributions, the final-spin-summed probability for a quark from nucleus A at a given location $(x_\mathrm{A},y_\mathrm{A})$ to scatter with the entire nucleus B can be written as 
\begin{equation}
\label{eq:sumAverageB}
\mathcal{P}_\mathrm{A} (x_\mathrm{A},y_\mathrm{A}) = \int T_2(x_\mathrm{B},y_\mathrm{B}) \frac{d \sigma(x_\mathrm{A},y_\mathrm{A},x_\mathrm{B},y_\mathrm{B})}{dx_\mathrm{B}dy_\mathrm{B}} dx_\mathrm{B}dy_\mathrm{B},
\end{equation}
where an integral weighted by the participant density in B is implemented. The same definition can also be applied for the  spin difference in the scattering probability $\Delta \mathcal{P}_\mathrm{A}(x_\mathrm{A},y_\mathrm{A})$ from its corresponding cross section $d \Delta \sigma$. Similarly, one can also obtain the spin-summed and spin-different probability -- $\mathcal{P}_\mathrm{B}(x_\mathrm{B},y_\mathrm{B})$ and $\Delta \mathcal{P}_\mathrm{B}(x_\mathrm{B},y_\mathrm{B})$ -- for a quark in nucleus B at $(x_\mathrm{B},y_\mathrm{B})$ to scatter with the entire nucleus A. By combining contributions from A and B, we obtain the following polarization of nuclear matter at a given location:
\begin{equation}
\label{eq:localInitialP}
P(x,y) = \frac{T_1(x,y) \Delta \mathcal{P}_\mathrm{A}(x,y) + T_2(x,y) \Delta \mathcal{P}_\mathrm{B}(x,y) } {T_1(x,y)\mathcal{P}_\mathrm{A}(x,y) + T_2(x,y) \mathcal{P}_\mathrm{B}(x,y) }.
\end{equation} 

\begin{figure}[tbp!]
	\begin{center}
	\includegraphics[width=0.4\textwidth]{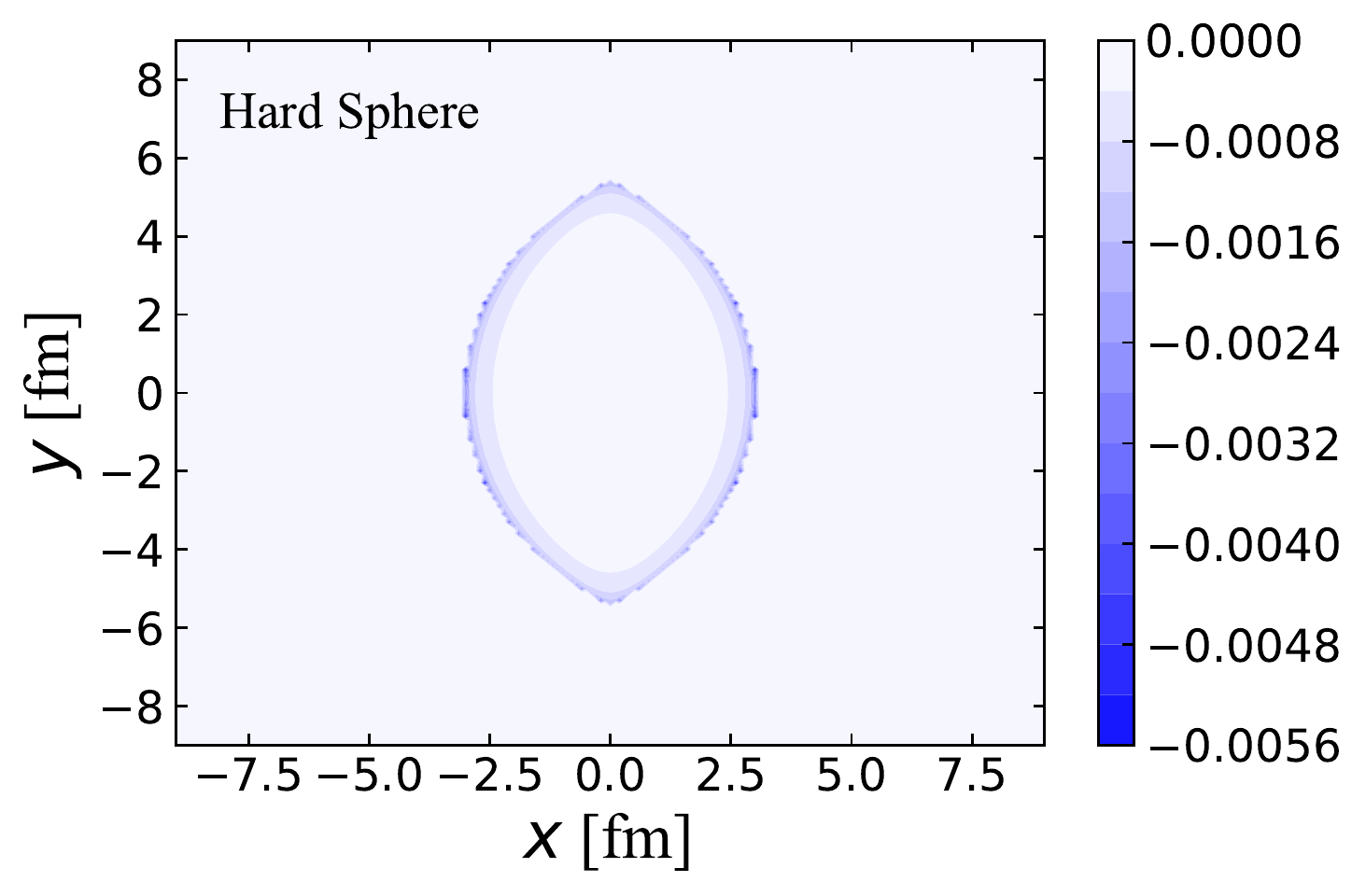}		
	\includegraphics[width=0.4\textwidth]{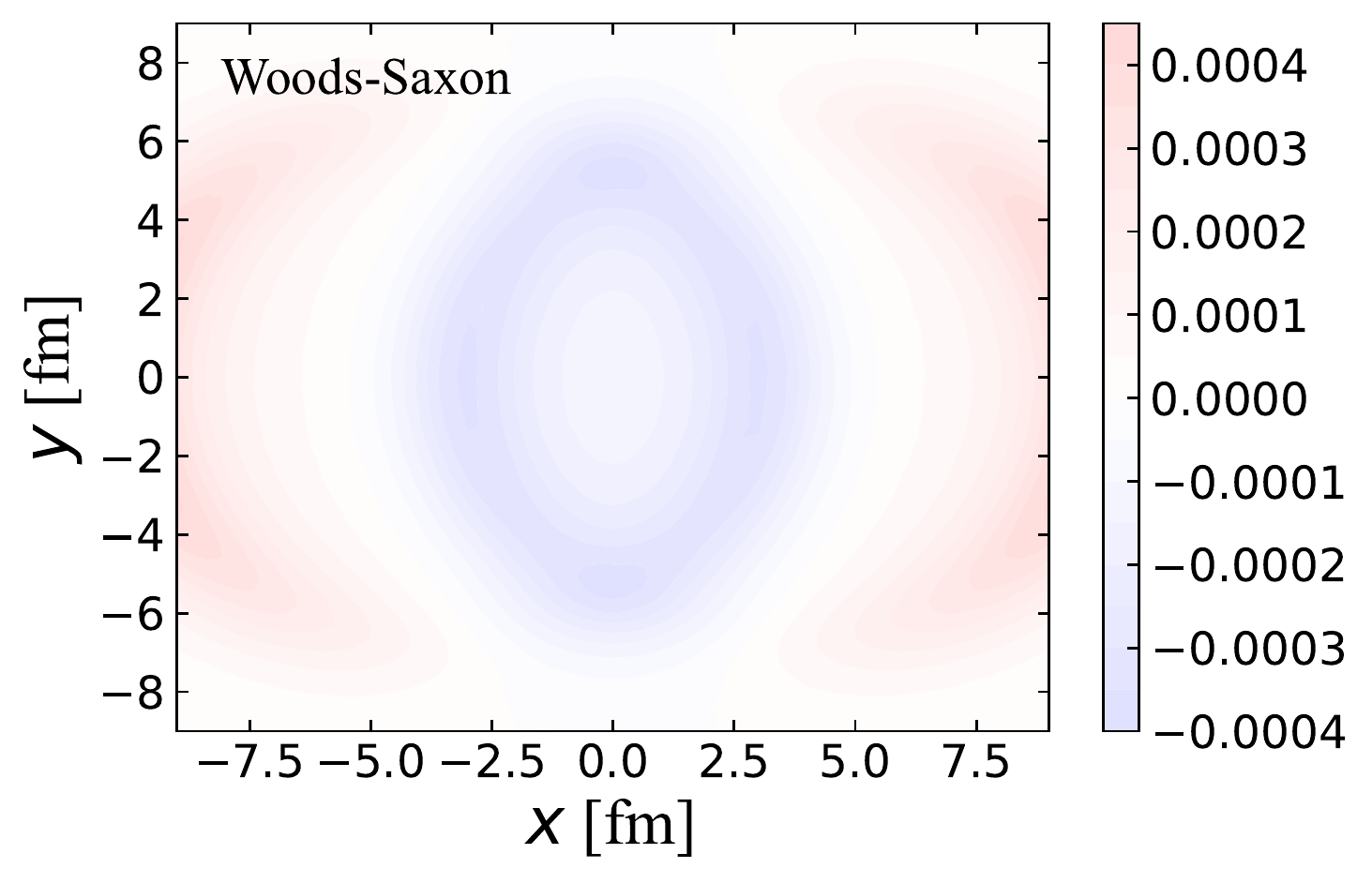}	
	\end{center}	
	\caption{(Color online) Initial spatial distribution of polarization in Au-Au collisions at $\sqrt{s_\mathrm{NN}}=200$~GeV with impact parameter $b=6.7$~fm, compared between Hard Sphere (upper panel) and Woods-Saxon (lower panel) distributions of nuclear density.}
	\label{fb}
\end{figure}

\begin{figure}[tbp!]
	\begin{center}
		\includegraphics[width=0.35\textwidth]{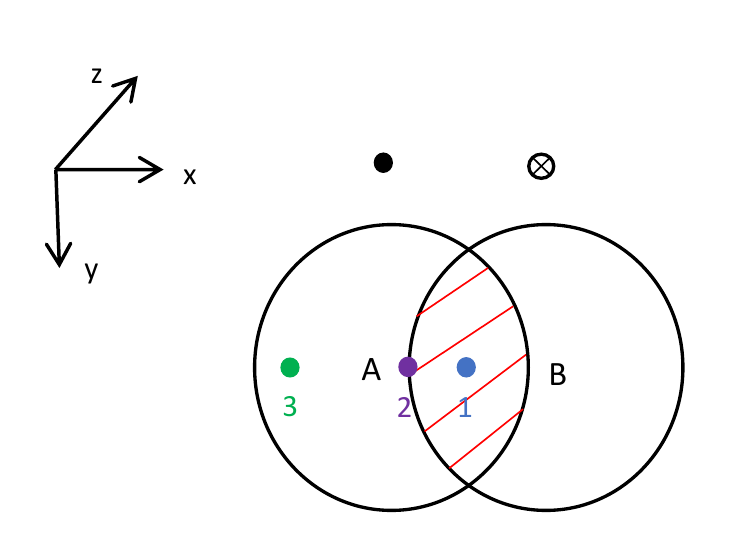}		
	\end{center}	
	\caption{(Color online) Illustration of quark scatterings at different locations.}
	\label{differentLocation}
\end{figure}

In Fig.~\ref{fb}, we first present the spatial distribution of the initial polarization calculated via Eq.~(\ref{eq:localInitialP}) above. A Au-Au collision system at $\sqrt{s_\mathrm{NN}}=200$~GeV is used here, with the impact parameter set as $b=6.7$~fm, approximating the 5-40\% centrality region. Comparing between the two subfigures, one can observe the dependence of polarization on the nuclear distribution function. In the upper panel, we find large polarization values mainly distribute around the boundary of the overlapping region between the two nuclei for the Hard Sphere distribution. This could be understood with the illustration in Fig.~\ref{differentLocation}. For location 1 that is at the center of the overlapping region, a quark from A interacts with similar number of participants from B on its left ($x<0$) and right ($x>0$), cancelling polarization due to our earlier discussion for Eq.~(\ref{eq:1xsectionDiff0}). On the other hand, for location 2 that resides on the boundary, a quark from A ``sees" most participants from B on its right, contributing to a negative polarization. Although the opposite conclusion (positive polarization) is drawn for a quark from B at location 2, its density [$T_2$ in Eq.~(\ref{eq:localInitialP})] is much smaller than that from A [$T_1$] at this location, leading to a net negative value of polarization in the end. Since there is no interaction outside the overlapping region within the Hard Sphere model, polarization is also zero in the corresponding region. Different distribution of polarization can be observed in the lower panel for the Woods-Saxon model. While the polarization is also small at the center and maximized around the boundary of the overlapping region, its magnitude decreases smoothly outside the boundary instead of suddenly vanishes because of the different location $(x,y)$ dependences of the nuclear density between Eq.~(\ref{eq:thicknessHS}) and~Eq.~(\ref{eq:thicknessWS}). It is interesting to note that for the case of Woods-Saxon model, one can also observe positive polarization value far away from the center. As illustrated in Fig.~\ref{differentLocation}, at location 3, a quark from A is polarized along $-\hat{y}$ by interacting with nucleus B; while a quark from B is polarized along $\hat{y}$ by interacting with A. Since the scattering cross section Eqs.~(\ref{eq:1xsectionTotal0}) and~(\ref{eq:1xsectionDiff0}) rapidly decreases with distance, the magnitude of the above polarization for the quark in A [$\Delta\mathcal{P}_\mathrm{A}$ in Eq.~(\ref{eq:localInitialP}) from Eq.~(\ref{eq:sumAverageB})] is much smaller than that of the quark in B [$\Delta\mathcal{P}_\mathrm{B}$], leading to a net positive polarization in the end. However, since the participant number densities [both $T_1$ and $T_2$ in Eq.~(\ref{eq:localInitialP})] are small at location 3, this positive value has little contribution to the global polarization after we integrate over the entire transverse plane, as will be shown later in this work.

\begin{figure}[tbp!]
	\centering
	\includegraphics[scale=0.35]{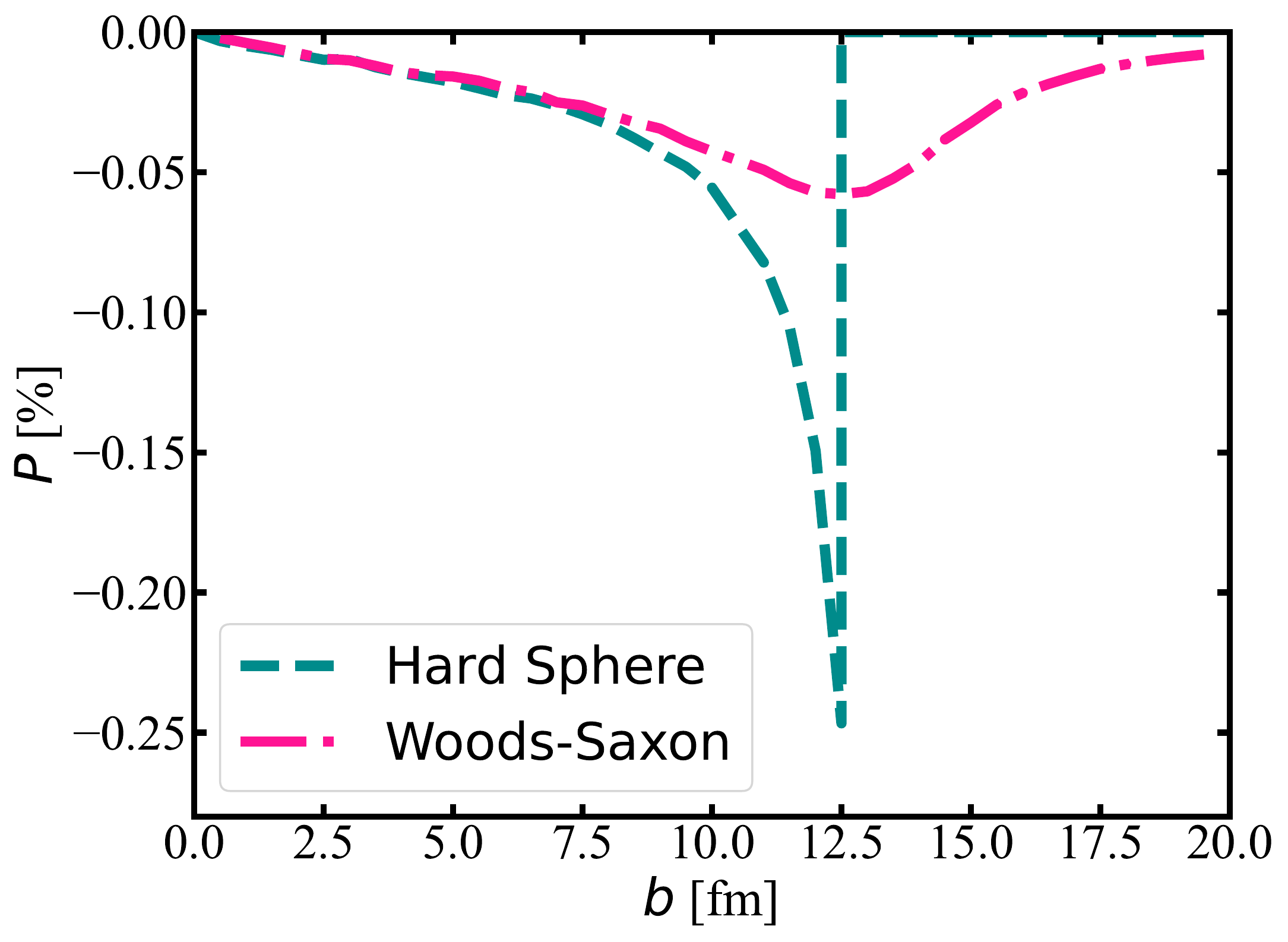}
	\addtocounter{figure}{0}
	\caption{(Color online) The initial global polarization as a function of the impact parameter, compared between  Hard Sphere and Woods-Saxon nuclear density distributions.}
	\label{fig:initialP}
\end{figure}

By integrating over the transverse plane, we may define the global polarization as follows,
\begin{align}
\label{eq:globalInitialP}
\overline{P}  =  \frac{\int \left[T_1(x,y) \Delta \mathcal{P} _\mathrm{A}(x,y) +T_2(x,y) \Delta \mathcal{P} _\mathrm{B}(x,y) \right] dx dy }{\int \left[T_1(x,y) \mathcal{P}_\mathrm{A}(x,y) +T_2(x,y)  \mathcal{P} _\mathrm{B}(x,y) \right] dx dy }.
\end{align}
The corresponding value is presented in \fig\ref{fig:initialP} as a function of the impact parameter for Au-Au collisions at $\sqrt{s_\mathrm{NN}}=200$~GeV. As expected, the global polarization is along the $-\hat{y}$ direction, reflected by its negative value, because on average participants from B is on the right ($x>0$) of participants from A. As the impact parameter increases, the magnitude of polarization first increases due to more asymmetric collisions, but then decreases due to vanishing participant nucleons in the initial hard scatterings. The maximum magnitude is obtained when $b$ is around $2R$ for sideswipe between the two nuclei. Beyond that, the polarization suddenly disappears for the Hard Sphere model due to the hard cut-off of its nuclear overlap function, while smoothly decreases to zero for the Woods-Saxon model. Results from these two nuclear density distributions are almost identical to each other for small impact parameter ($b\lesssim R$); noticeable difference is only observed at larger $b$. In this work, we initialize polarization according to smooth nuclear density distributions. Effect from event-by-event fluctuations on the global polarization was shown small in Ref.~\cite{Karpenko:2021wdm}. On the other hand, the local polarization along the longitudinal direction due to the radial flow of the QGP might be more sensitive to the initial state fluctuation. This will be explored in our future work by implementing our current framework using the Monte-Carlo method.

%
%

\section{Evolution of polarization through the QGP phase}
\label{sec3}
\subsection{Polarization of initially polarized quarks}
\label{subsec:rescattering}

In the previous section, we discussed the production of global polarization from initial hard collisions between nuclei, where Eqs.~(\ref{eq:xsection0})-(\ref{eq:1xsectionDiff0}) were derived for initially unpolarized quarks. To investigate the further evolution of polarization through the QGP phase, one needs to extend these equations to the scenario where the projectile quark already possesses non-zero polarization. This has been developed in Ref.~\cite{Huang:2011ru}, where Eq.~({\ref{eq:xsection0}}) is re-written as 
\begin{equation}
\begin{aligned}	
		\frac{d\sigma_{\lambda_f}}{d^2x_\mathrm{T}}=\;& C_\mathrm{T}\sum_{\lambda_i}\int\frac{d^2q_\mathrm{T}}{(2\pi)^2}\frac{d^2k_\mathrm{T}}{(2\pi)^2} e^{i(\vec{k}_\mathrm{T}-\vec{q}_\mathrm{T})\cdot\vec{x}_\mathrm{T}}\\
		& \times R_{\lambda_i} \mathcal{I}_{\lambda_f \lambda_i}(\vec{k}_\mathrm{T},\vec{q}_\mathrm{T},E),
\end{aligned}	
\end{equation}	
in which $R_{\lambda_i} = (1+\lambda_i P_i)/2$, with $P_i$ being the initial polarization of the projectile quark. Using the small scattering angle approximation as before, the $\mathcal{I}_{\lambda_f \lambda_i}$ part can be approximated with
\begin{align}
\label{eq:I1}
\frac{\mathcal{I}_{\lambda_f \lambda_i}}{g^2} & \approx \frac{1}{2}A_0(q_\mathrm{T})A_0(k_\mathrm{T})\Bigg[ 1+\lambda_i \lambda_f \nonumber\\
&-i(\lambda_i+\lambda_f)\frac{(\vec{q}_\mathrm{T}-\vec{k}_\mathrm{T})\cdot(\hat{n}\times\vec{p})}{2E(E+m)}\Bigg],
\end{align}
which returns to Eq.~(\ref{eq:I0simplify}) after the initial spin states ($\lambda_i$) are averaged over.
The differential cross section is then simplified to
\begin{align}
		\frac{d\sigma_{\lambda_f}}{d^2x_\mathrm{T}}&= \frac{1}{2}g^4C_\mathrm{T} \Bigg[\frac{1}{4\pi^2}(1+\lambda_f P_i)K^2_0(\mu x_\mathrm{T}) \\
		&- \frac{\mu}{4\pi^2}\frac{(\lambda_f+P_i) \vec{p} \cdot ( \hat{x}_\mathrm{T} \times \hat{n} )}{E(E+m)} K_0(\mu x_\mathrm{T})K_1(\mu x_\mathrm{T})\Bigg].\nonumber
\end{align}
The final-spin-independent sum of the cross section reads
\begin{equation}
\label{eq:xsectionTotal0}
	\begin{aligned}
		\frac{d \sigma}{d^2x_\mathrm{T}}&=\frac{d\sigma_+}{d^2x_\mathrm{T}} + \frac{d\sigma_-}{d^2x_\mathrm{T}} \\
		&= 4\alpha_\mathrm{s}^2 C_\mathrm{T}\Bigg[ K^2_0(\mu x_\mathrm{T}) \\
		& - \mu P_i  \frac{\vec{p} \cdot (\hat{x}_\mathrm{T}\times \hat{n}) }{E(E+m)}K_0(\mu x_\mathrm{T})K_1(\mu x_\mathrm{T}) \Bigg],		
	\end{aligned}
\end{equation}
while the final-spin-dependent difference reads
\begin{equation}
\label{eq:xsectionDiff0}
	\begin{aligned}
		\frac{d\Delta \sigma}{d^2x_\mathrm{T}}&=\frac{d\sigma_+}{d^2x_\mathrm{T}} -\frac{d\sigma_-}{d^2x_\mathrm{T}}\\
		&= 4\alpha_\mathrm{s}^2 C_\mathrm{T}\Bigg[ P_i K^2_0(\mu x_\mathrm{T}) \\
		&- \mu \frac{ \vec{p} \cdot (\hat{x}_\mathrm{T}\times \hat{n}) }{E(E+m)} K_0(\mu x_\mathrm{T}) K_1(\mu x_\mathrm{T}) \Bigg].	
	\end{aligned}
\end{equation}

Instead of integrating over the half $x$-$y$ plane for scattering centers always on one side of the projectile quark, as assumed in Ref.~\cite{Huang:2011ru}, we obtain the total cross section by integrating over the entire transverse plane, weighted by the local entropy density (or particle number density) of the QGP. Therefore, the final polarization of a quark residing at a given location $(x,y)$ is given by
\begin{align}
	P_f(x,y) = \frac{ \int dx_1 dy_1 s(x_1,y_1) \frac{d\Delta \sigma(x,y,x_1,y_1) }{dx_1 dy_1}}{\int  dx_1 dy_1  s(x_1,y_1) \frac{d \sigma(x,y,x_1,y_1) }{dx_1 dy_1 }  },
\end{align}
where the local entropy density $s$ can be taken from a hydrodynamic simulation of the QGP medium, as will be discussed in the next subsection.

The change of polarization after the scattering is then obtained as
\begin{align}
\label{eq:deltaP}
		& \Delta  P(x,y) = P_f(x,y)-P_i(x,y)\\ 
		&=  -\Bigg\{\int dx_1 dy_1 s(x_1,y_1) \Big[ \left(1-P^2_i(x,y)\right) \frac{ \mu \vec{p} \cdot (\hat{x}_\mathrm{T} \times \hat{n}) }{E(E+m)} \nonumber\\
		& \times K_0(\mu x_\mathrm{T}) K_1(\mu x_\mathrm{T}) \Big]\Bigg\}  \Bigg/ \Bigg\{ \int dx_1 dy_1 s(x_1,y_1)\times\nonumber  \\
		& \Big[K^2_0(\mu x_\mathrm{T}) - P_i(x,y) \frac{ \mu \vec{p} \cdot (\hat{x}_\mathrm{T} \times \hat{n}) }{E(E+m)} K_0(\mu x_\mathrm{T})K_1(\mu x_\mathrm{T})\Big]\Bigg\}.\nonumber
\end{align}

%
%

For multiple scatterings inside the QGP medium, the relative momentum $\vec{p}$ in Eq.~(\ref{eq:deltaP}) is taken as $E v_z$ for estimating the polarization along the $\hat{y}$ direction, where $E=\epsilon/\rho$ represents the energy of a quark with $\epsilon$ and $\rho$ being the local energy and particle number densities inside the QGP, and $v_z$ is the local fluid velocity along the $\hat{z}$ direction. In addition, we assume the mean free path of a quark is $\tau_q$, which can be roughly related to the shear viscosity of the QGP via $\eta_v \approx (1/3) \rho \langle p_\mathrm{th} \rangle (4/9) \tau_q \approx (4/9) T \rho \tau_q$ for a thermal ensemble of gluons~\cite{Danielewicz:1984ww,Huang:2011ru}, with $\langle p_\mathrm{th} \rangle = 3T$ being the average thermal momentum at temperature $T$. This leads to the following equation for the time evolution of polarization:
\begin{align}
\label{eq:evoEq}
	\frac{dP(x,y)}{dt} = \frac{\Delta P(x,y)}{\tau_q} = \frac{4T\rho }{9s} \frac{s}{\eta_v} \Delta P(x,y).
\end{align}
We will use this equation to evolve the quark polarization at each location $(x,y)$ inside the QGP. The local temperature ($T$), number density ($\rho$) and entropy density ($s$) can be provided by the hydrodynamic model, and the shear-viscosity-to-entropy-density ratio is taken as $\eta_v/s=0.08$.

\subsection{Hydrodynamic simulation of the QGP}
\label{eq:hydro}

In this work, we use the (3+1)-dimensional viscous hydrodynamic model CLVisc~\cite{Pang:2018zzo,Wu:2018cpc} to simulate the spacetime evolution of the QGP medium. Following our earlier work~\cite{Jiang:2021ajc}, three different model setups are used to generate the initial energy density distribution of the medium, which is fed as the initial condition into the hydrodynamic evolution. By comparing between these three model setups, one may explore how the final-state global polarization depends on the initial geometry of the medium.

\underline{\textbf{\emph{Case (A)}}} Boz$\dot{\textrm{e}}$k-Wyskiel parametrization.

\begin{figure}[tbp!]
\begin{center}
\includegraphics[width=0.4\textwidth]{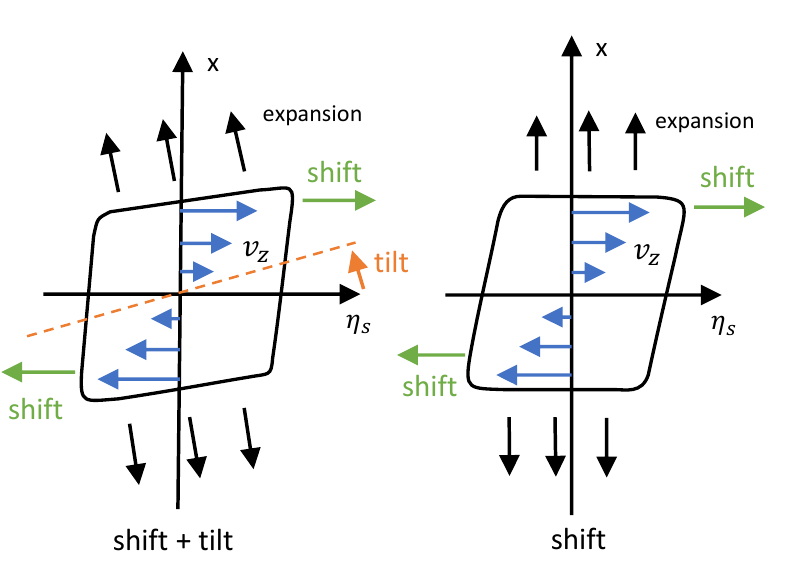}		
\end{center}	
\caption{(Color online) Illustration of the initial energy density distribution on the $x$-$\eta_\mathrm{s}$ plane, left for our Case (A) and Case (B) that include a counter-clockwise tilt, and right for our Case (C) that only includes a longitudinal shift.}
	\label{fig:initialDeformation}
\end{figure}

Our first parametrization of the initial condition is based on Refs.~\cite{Bozek:2010bi,Bozek:2011ua}, which is one of the earliest and most widely applied initialization scheme that takes into account the deformation of nuclear matter in the reaction plane due to non-central collisions, as illustrated in Fig.~\ref{fig:initialDeformation}. In this model, the transverse ($x$, $y$) and longitudinal (spacetime rapidity $\eta_\mathrm{s}$) distribution of the energy density is parametrized with a wounded nucleon weight function as follows,
\begin{equation}
\label{eq:}
W_\mathrm{N}(x,y,\eta_\mathrm{s}) = 2[T_2(x,y)f_+(\eta_\mathrm{s}) + T_1(x,y)f_-(\eta_\mathrm{s})], 
\end{equation}
in which $T_{2,1}$ is the participant nucleon density distributions of the $\pm \hat{z}$-going nucleus, as discussed in Eq.~(\ref{eq:partWS}). The Woods-Saxon model of the nucleon distribution is used for initializing the hydrodynamic evolution. The $f_{\pm}$ function is designed to introduce the geometric asymmetry along the longitudinal direction as 
$$ f_{+}(\eta_{s})=\left\{
\begin{aligned}
&0,                                 && \eta_\mathrm{s} < -\eta_\mathrm{m}, \\
&\frac{\eta_\mathrm{s}+\eta_\mathrm{m}}{2\eta_\mathrm{m}},  && -\eta_\mathrm{m} \leq \eta_\mathrm{s} \leq\eta_\mathrm{m},\\
&1,                                 & &\eta_\mathrm{s} > \eta_\mathrm{m},
\end{aligned}
\right.
$$
and
$$ f_{-}(\eta_\mathrm{s})=\left\{
\begin{aligned}
&1,                                 && \eta_\mathrm{s} < -\eta_\mathrm{m}, \\
&\frac{-\eta_\mathrm{s}+\eta_\mathrm{m}}{2\eta_\mathrm{m}},  && -\eta_\mathrm{m} \leq \eta_\mathrm{s} \leq\eta_\mathrm{m},\\
&0,                                 & &\eta_\mathrm{s} > \eta_\mathrm{m},
\end{aligned}
\right.
$$
where $\eta_\mathrm{m}$ defines the range of rapidity correlations, which affects the relative contribution from forward and backward participating nucleons. This parametrization introduces not only a longitudinal shift but also a counter-clockwise tilt in the $x$-$\eta_\mathrm{s}$ plane to the 3-dimensional geometry of the medium, as illustrated in the left panel of Fig.~\ref{fig:initialDeformation}.

The full expression of the initial energy density distribution is then given by
\begin{equation}
\label{eq:energyDensityBW}
	\epsilon (x,y,\eta_\mathrm{s}) = K \cdot W(x,y,\eta_\mathrm{s}) \cdot H(\eta_\mathrm{s}),
\end{equation}
where $K$ is an overall normalization factor that is fitted to the final charged particle yield ($dN_\mathrm{ch}/d\eta$ with $\eta$ being the pseudorapidity) observed in relativistic heavy-ion collisions. A function $H(\eta_\mathrm{s})$
\begin{equation}
H(\eta_\mathrm{s}) = \exp \left[ -\frac{(|\eta_\mathrm{s}| - \eta_\mathrm{w})^2}{2\sigma ^2_{\eta}} \theta (|\eta_\mathrm{s}| - \eta_\mathrm{w}) \right]
\end{equation}
is introduced to describe the plateau pattern of $dN_\mathrm{ch}/d\eta$ with respect to $\eta$, with $\eta_\mathrm{w}$ and $\sigma_\eta$ as two model parameters. The total weight function 
\begin{equation}
W(x,y,\eta_\mathrm{s}) = \frac{(1-\alpha ) W_\mathrm{N}(x,y,\eta_\mathrm{s}) + \alpha n_\mathrm{BC}(x,y) }{[(1-\alpha)W_\mathrm{N}(0,0,0) + \alpha n_\mathrm{BC}(0,0)]|_{b=0}},\end{equation}
combines contributions from wounded nucleons and binary collisions -- the latter is given by
\begin{equation}
\begin{aligned}
n_\text{BC}(x,y)=\sigma_\text{NN}T_\mathrm{A}(x,y)T_\mathrm{B}(x,y).
\label{eq:nbc}
\end{aligned}
\end{equation}
The parameter $\alpha$ determines the relative contribution from wounded nucleons and binary collisions, which can be extracted from the impact parameter dependence of $dN_\mathrm{ch}/d\eta$. Related model parameters will be listed later in this subsection when we calculate for a given collision system.

\underline{\textbf{\emph{Case (B)}}} CCNU parametrization.

An alternative parameterization of the deformed initial energy density was developed in Ref.~\cite{Jiang:2021foj}, which is similar to the above Boz$\dot{\textrm{e}}$k-Wyskiel ansatz -- Eqs.~(\ref{eq:energyDensityBW}) - (\ref{eq:nbc}), except that the longitudinal dependence of the wounded nucleon weight function is parametrized as 
\begin{equation}
\label{eq:wCCNU}
\begin{aligned}
W_\mathrm{N}(x,y,\eta_\mathrm{s}) & = [T_1(x,y) + T_2(x,y)] \\ 
& + H_\mathrm{t} [T_2(x,y) - T_1(x,y)] \tan \left( \frac{\eta_\mathrm{s}}{\eta_\mathrm{t}} \right).
\end{aligned}
\end{equation}
Two parameters -- $H_\mathrm{t}$ and $\eta_\mathrm{t}$ -- are introduced to describe the unbalanced energy deposition, between the projectile and target nuclei at different transverse locations, into the medium at different spacetime rapidities. Similar to the Boz$\dot{\textrm{e}}$k-Wyskiel setup, this CCNU parameterization will also cause both shift and tilt of the initial energy density distribution as illustrated in the left panel of Fig.~\ref{fig:initialDeformation}.

\underline{\textbf{\emph{Case (C)}}} Shen-Alzhrani parametrization.

The third $\eta_\mathrm{s}$-dependent initial condition model was adopted from Refs.~\cite{Shen:2020jwv,Ryu:2021lnx}, which ensures the local energy-momentum conservation when convert the two colliding nuclei into the energy density profile of the hot nuclear medium. One first defines the local invariant mass $M(x,y)$ and the center-of-mass rapidity $y_{\text{CM}}$ as,
\begin{equation}
M(x,y) = m_\text{N}\sqrt{T_{1}^{2}+T_{2}^{2}+2T_{1}T_{2}\textrm{cosh}(2y_{\textrm{beam}})},
\label{eq:mxy}
\end{equation}
\begin{equation}
y_{\textrm{CM}}(x,y) = \textrm{arctanh}\left[\frac{T_{2}-T_{1}}{T_{1}+T_{2}}\textrm{tanh}(y_{\textrm{beam}})\right],
\label{eq:ycm}
\end{equation}
where $y_{\textrm{beam}}=\textrm{arccosh}(\sqrt{s_\text{NN}}/2m_\text{N})$ is the rapidity of each nucleon inside the colliding beams, with $m_\text{N}$ being its mass.

The initial energy density profile is then constructed as~\cite{Shen:2020jwv},
\begin{align}
\varepsilon(x,y,\eta_\mathrm{s};&\; y_\mathrm{CM})=K \cdot\mathcal{N}_{e}(x,y)\nonumber\\
&\times \exp{\Big [}-\frac{(|\eta_\mathrm{s}-(y_\mathrm{CM}-y_\mathrm{L})|-\eta_\mathrm{w})^{2}}{2\sigma^{2}_{\eta}}\nonumber \\
&\times \theta(|\eta_\mathrm{s}-(y_\mathrm{CM}-y_\mathrm{L})|-\eta_\mathrm{w}){\Big]}.
\label{eq:eqosu1}
\end{align}
Same as the previous two models, $K$ is the overall normalization factor, and $\eta_\mathrm{w}$ and $\sigma_{\eta}$ are the width parameters for the plateau width of $dN_\mathrm{ch}/d\eta$ distribution with respect to $\eta$. A new parameter --  $y_\mathrm{L}=f y_\mathrm{CM}$ with $f \in [0,1]$ -- is introduced in this model to describe the deformation of the medium along the longitudinal direction. The transverse density distribution $\mathcal{N}_{e}$ is determined by the local invariant mass $M(x,y)$ as
\begin{equation}
\begin{aligned}
\mathcal{N}_{e}(x,y) =  \frac{M(x,y)}{{\color{black}M(0,0)}\left[2\sinh(\eta_\mathrm{w})+\sqrt{\frac{\pi}{2}}\sigma_{\eta}e^{\sigma^{2}_{\eta}/2}C_{\eta}\right]},
\label{eq:Nxy}
\end{aligned}
\end{equation}
\begin{equation}
\begin{aligned}
C_{\eta} = e^{\eta_\mathrm{w}}\textrm{erfc}\left(-\sqrt{\frac{1}{2}}\sigma_{\eta}\right)+e^{-\eta_\mathrm{w}}\textrm{erfc}\left(\sqrt{\frac{1}{2}}\sigma_{\eta}\right),
\end{aligned}
\end{equation}
where erfc($x$) is the complementary error function. Different from Boz$\dot{\textrm{e}}$k-Wyskiel and CCNU parameterizations, this Shen-Alzhrani parametrization only generates a shift deformation along the longitudinal direction, as illustrated in the right panel of Fig.~\ref{fig:initialDeformation}; tilt of the medium profile has not been included yet. A detailed comparison of this deformation can be found in our earlier work~\cite{Jiang:2021ajc}.

In many hydrodynamic calculations, including our previous study~\cite{Jiang:2021ajc}, local flow velocities are initialized with zero. While this simplification has minor effect on observables that are mainly driven by the QGP expansion, e.g. the yield of charged particles and their harmonic flow coefficients, it ignores the initial orbital angular momentum deposited into the system and thus would fail in describing observables related to the global polarization. As revealed in Ref.~\cite{Huang:2011ru}, the velocity gradient $\partial v_z/\partial x$ is the main origin of the global polarization generated inside the QGP. Therefore, for all cases (A, B and C above), we follow Ref.~\cite{Shen:2020jwv} to initialize the off-diagonal components of the energy-momentum tensor as
\begin{align}
	T^{\tau \tau}(x,y,\eta_\mathrm{s}) &= \epsilon(x,y,\eta_\mathrm{s}) \cosh(y_\mathrm{L}), \\
	T^{\tau \eta_\text{s}}(x,y,\eta_\mathrm{s}) &= \frac{1}{\tau_0}\epsilon(x,y,\eta_\mathrm{s}) \sinh(y_\mathrm{L}).
\end{align}
The initial flow velocity in the longitudinal direction is then given by their ratio as
\begin{equation}
	v_{\eta_\mathrm{s}}=T^{\tau \eta_\text{s}}/T^{\tau \tau}.
\end{equation}
From the above equations, one notices that the $y_\mathrm{L}$ (or $f$) parameter designed in Case (C) determines the amount of the longitudinal momentum from the beam nucleons that is deposited into the QGP medium as its initial longitudinal velocity. In the present study, we only introduce the non-zero initialization of the longitudinal velocity, the transverse components of the energy-momentum tensor ($T^{\tau x}$ and $T^{\tau y}$), or the corresponding flow velocities ($v_x$ and $v_y$) are still initialized as zero, since they are not expected to affect the global polarization which aligns with the $\hat{y}$ direction. 

Based on the above initial energy density and flow velocity, we then apply the CLVisc hydrodynamic model~\cite{Pang:2018zzo,Wu:2018cpc} to simulate the subsequent spacetime evolution of the QGP profiles, starting from an initial proper time $\tau_0$. The hydrodynamic equations read:
\begin{align}
	\partial _{\mu} T^{\mu \nu} &=0,
\end{align}
where the energy-momentum tensor is given by
\begin{align}
	T^{\mu \nu} & = \epsilon u^{\mu} u^{\nu} - (P+\Pi) \Delta ^{\mu \nu} + \pi ^{\mu \nu}.
\end{align}
Here, $\epsilon$ is the local energy density, $u^{\mu}$ is the fluid four-velocity, $P$ is the pressure, $\pi ^{\mu \nu}$ is the shear stress tensor, and $\Pi$ is the bulk pressure (taken as zero in our current calculations); $g^{\mu \nu} =  \mathrm{diag}(1,-1,-1,-1)$ is the metric tensor and $\Delta ^{\mu \nu} = g^{\mu \nu} - u^{\mu} u^{\nu}$ is the projection tensor. These hydrodynamic equations are solved together with the lattice QCD Equation of State (EoS) from the Wuppertal-Budapest work (2014)~\cite{Borsanyi:2013bia}.

In this study, we apply the isothermal freeze-out conditions~\cite{Pang:2018zzo}, in which the freeze-out hypersurface is determined by a constant temperature $T_\mathrm{frz}$. Interaction between a projectile quark and its surrounding medium, and thus the evolution of its polarization, ceases when the local temperature drops below $T_\mathrm{frz}$.

In the rest of this paper, we will use the 5-40\% Au-Au collisions at $\sqrt{s_\mathrm{NN}}=200$~GeV, if not otherwise specified, as an example to study the evolution of quark polarization inside the QGP. In Tab.~{\ref{tab:parameters}}, we summarize all our model parameters mentioned earlier in this subsection, which were constrained in our previous study~\cite{Jiang:2021foj} by the soft hadron yield and their directed flow coefficient.
\begin{table}[!h]
\begin{center}
\begin{tabular}{|c|  c|  c|  c| c|  }
\hline \hline
$\tau_0$~[fm]  & $K$~[GeV/fm$^3$] & $\;\;\;\; \eta_{\mathrm{w}}\;\;\;\;$  &$\;\;\;\; \sigma_{\mathrm{\eta}} \;\;\;\;$   & $\;\;\;\; \alpha\;\;\;\;$   \\
\hline
0.6    &35.5      &1.3        &1.5     &0.05    \\
\hline \hline
$T_{\mathrm{frz}}$~[MeV]  &$\eta_\mathrm{m}$  &$H_{\mathrm{t}}$      & $\eta_\mathrm{t}$   &$f$ \\ 
\hline
137      &2.8     &2.9  &8.0   &0.15    \\
\hline \hline
\end{tabular}
\caption{\label{t:parameters} Model parameters for the hydrodynamic evolution~\cite{Jiang:2021foj} for Au-Au collisions at $\sqrt{s_{\mathrm{NN}}}=200$~GeV with impact parameter $b=6.7$~fm, among which $\eta_\mathrm{m}$ is for our Case (A), $H_{\mathrm{t}}$ and $\eta_\mathrm{t}$ are for Case (B), and the other parameters are commonly applied to all models here.}
\label{tab:parameters}
\end{center}
\end{table}

\begin{figure}[tbp]
\centering
\includegraphics[scale=0.35]{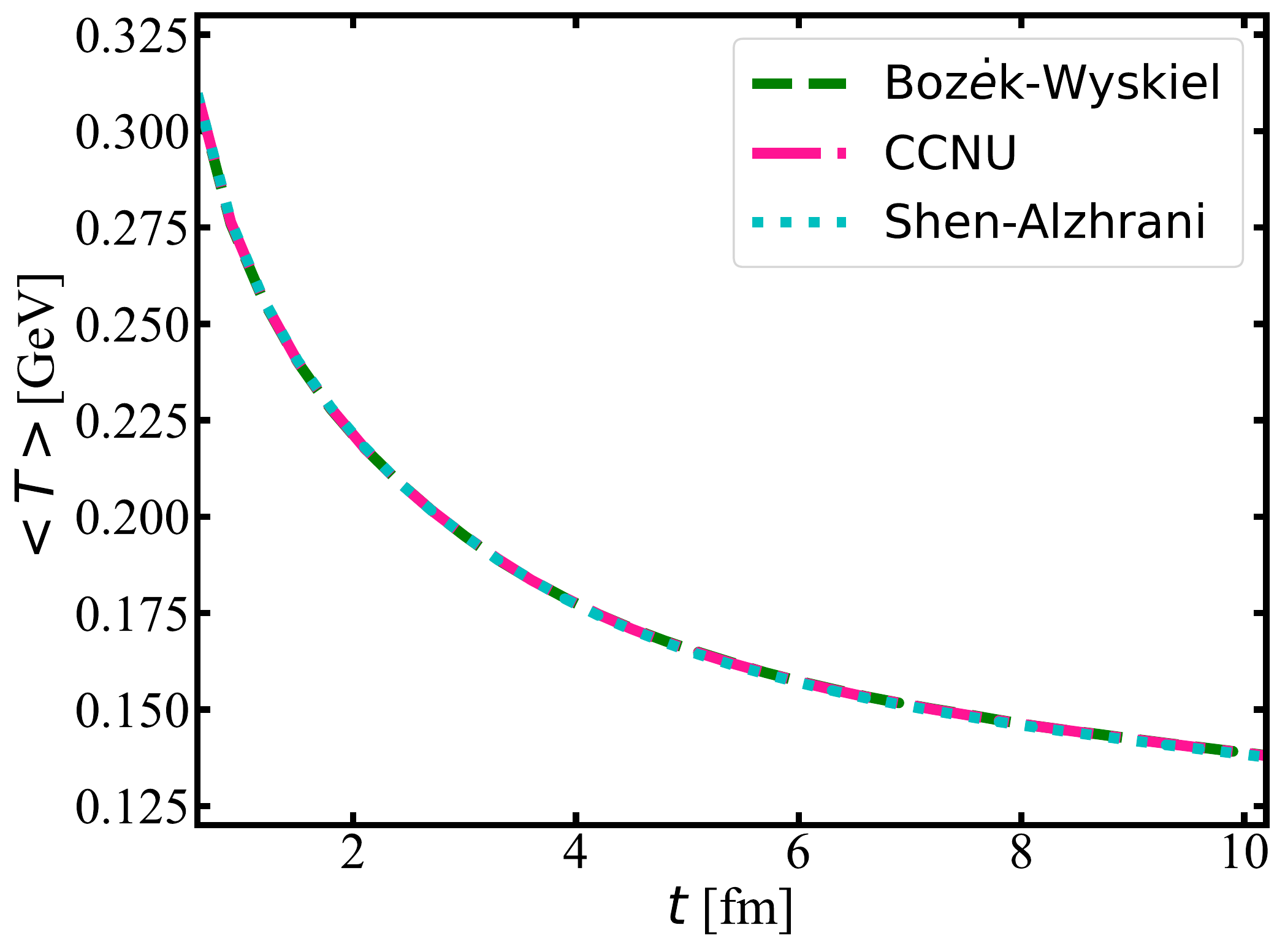}
\addtocounter{figure}{0}
\renewcommand{\figurename}{FIG}
\caption{(Color online) Time evolution of the average temperature $\left <T \right>$, compared between three different initial condition setups.}
\label{ta}
\end{figure}

As previously discussed for Eq.~(\ref{eq:deltaP}), the evolution of quark polarization inside the QGP depends on the medium temperature ($T$) and longitudinal velocity ($v_z$) profiles. Therefore, we first investigate how different initial conditions affect these quantities during the hydrodynamic expansion. In Fig.~\ref{ta}, we study the time evolution of the average temperature of the QGP fireball. The average is conducted on the $\eta_\mathrm{s} = 0$ (or $z=0$) plane over the hydrodynamic cells with local temperature above $T_\mathrm{frz}$. The local entropy density $s$ is applied as the weight for the average. As shown in Fig.~\ref{ta}, the average temperature decreases with time towards $T_\mathrm{frz}$. No visible difference can be observed for the temperature evolution between the three initial condition setups, because they are all adjusted to describe the soft hadron yield data in our earlier work~\cite{Jiang:2021foj}. 

%

\begin{figure}[tbp]
	\begin{center}
	\includegraphics[width=0.4\textwidth]{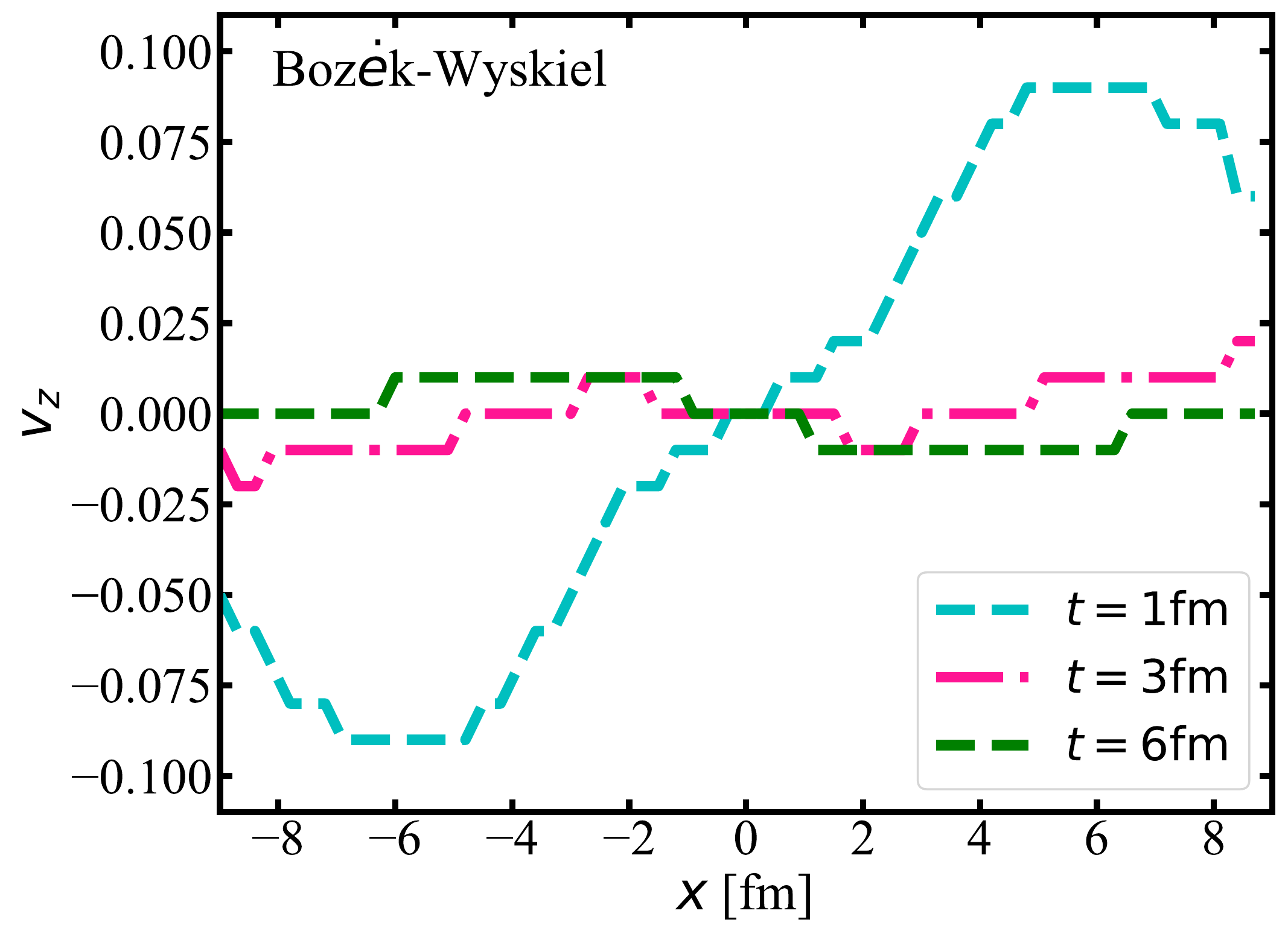}
	\includegraphics[width=0.4\textwidth]{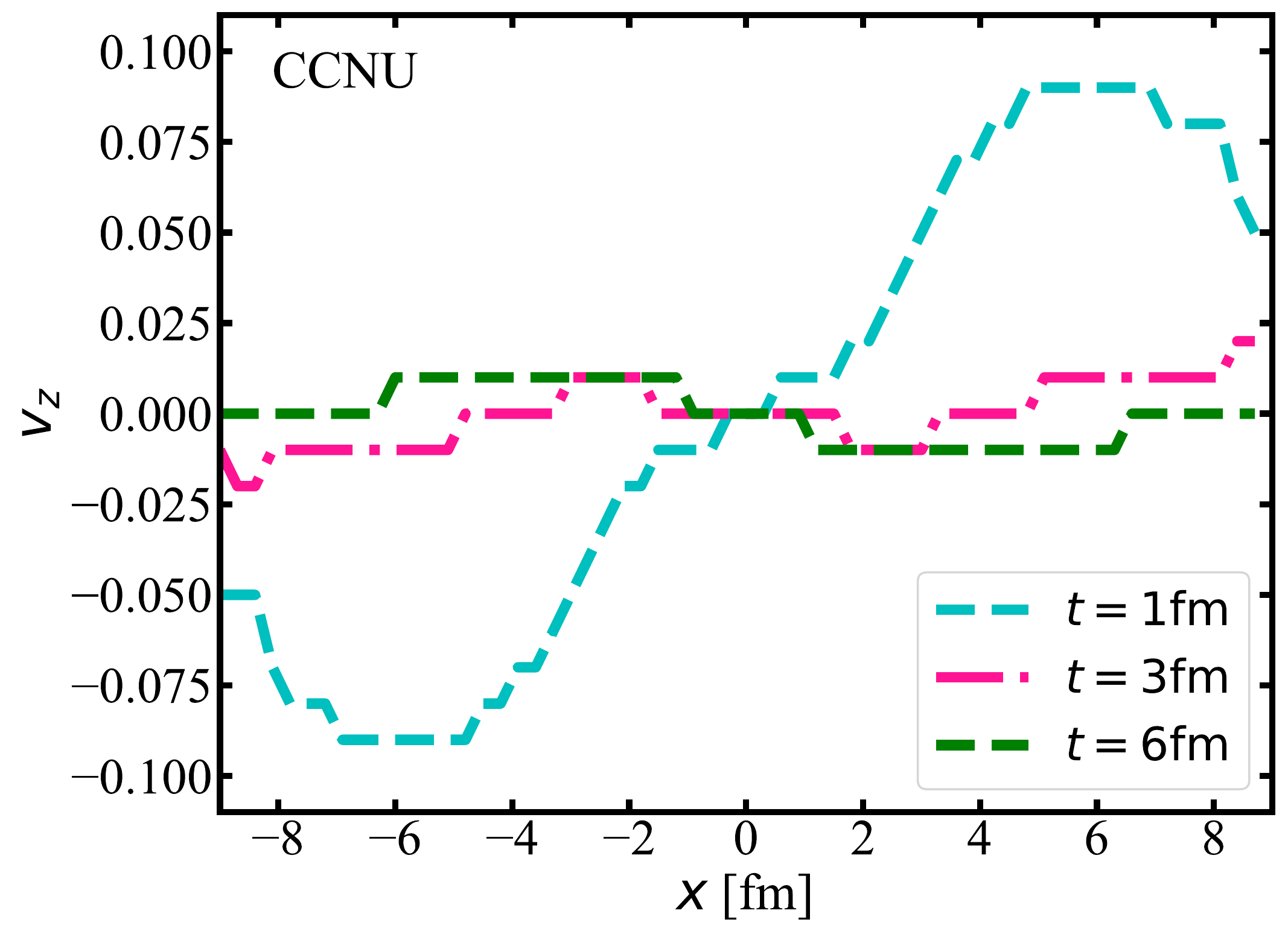}		
	\includegraphics[width=0.4\textwidth]{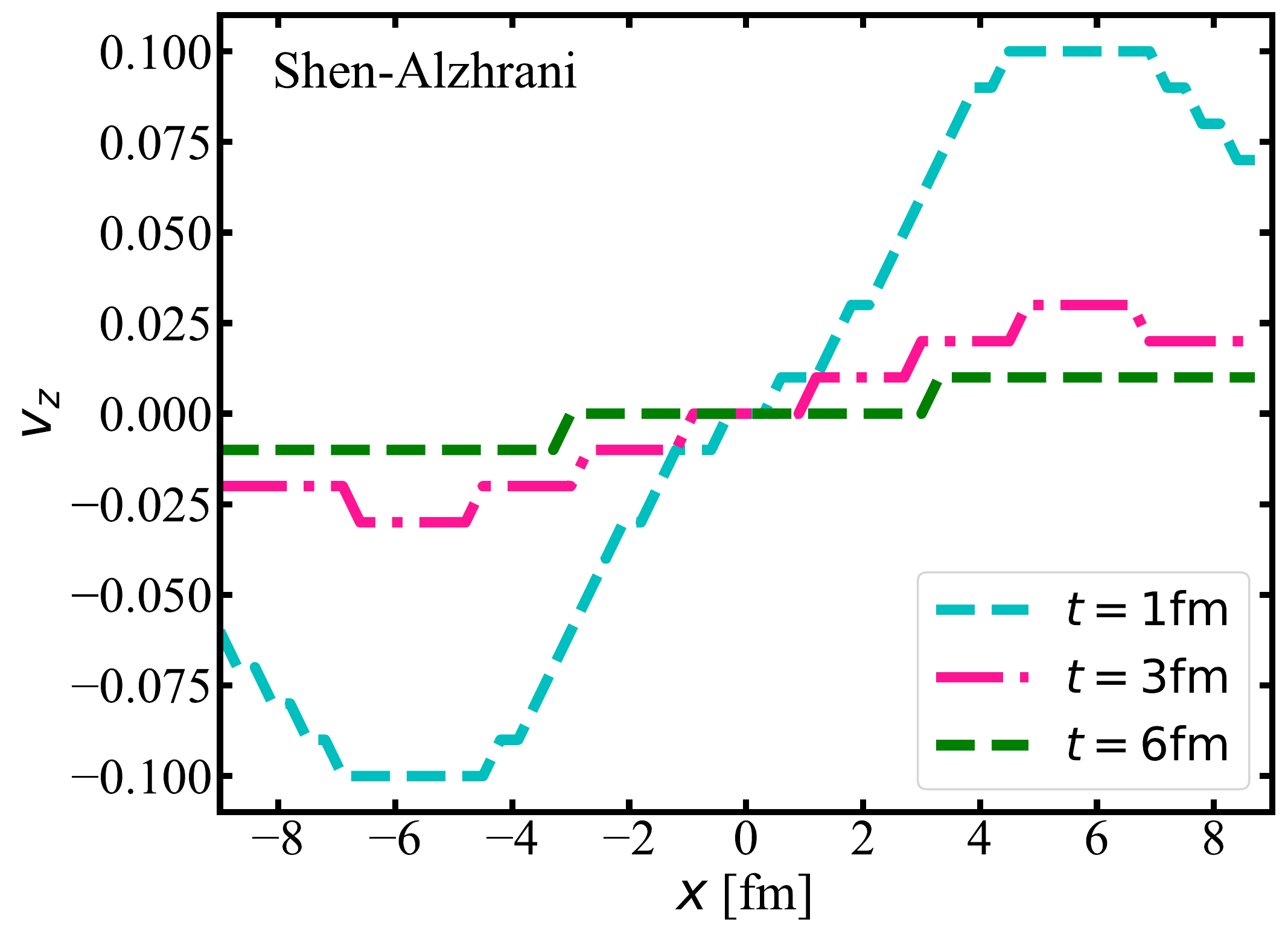}
	\end{center}
	\caption{(Color online) The longitudinal velocity distribution along the $x$-axis at different times. Results from three different initial condition setups are presented in the three panels.}
	\label{Vzx}
\end{figure}	

Shown in Fig.~\ref{Vzx} is the $v_z$ distribution along the $x$-axis ($y=z=0$) at different times. Due to the asymmetric collision, one expects to see $v_z>0$ in the $x>0$ half plane while $v_z<0$ in the $x<0$ half plane at the initial time. As time evolves, the magnitude of $v_z$ becomes smaller. It is interesting to note that while the magnitude of $v_z$ starting from the Shen-Alzhrani initial condition (bottom panel) decreases with time towards zero, the sign of $v_z$ from the other two initial conditions (top and middle panels) can flip at later time (e.g. $t=6$~fm), leading to  $v_z<0$ at $x>0$ and $v_z>0$ at $x<0$. This could be understood with the tilted geometry in the Boz$\dot{\textrm{e}}$k-Wyskiel and CCNU initial conditions as illustrated in the left panel of Fig.~\ref{fig:initialDeformation}, whose further expansion produces negative $v_z$ component in the $x>0$ plane, while positive $v_z$ component in the $x<0$ plane. This drives a quicker decay of $v_z$ from these two initial conditions (compared to the Shen-Alzhrani initial condition without such tilt) and in the end can also reverse the sign of $v_z$. Since the velocity gradient is the key origin of the development of quark polarization inside the QGP, we expect to obtain different magnitudes of the final-state global polarization from these initial conditions, as will be presented in the coming subsection. We also note that the magnitude of $v_z$ obtained from the realistic hydrodynamic simulation here is much smaller than the relativistic laminar flow model applied in the earlier study~\cite{Huang:2011ru}, which will affect the magnitude of global polarization for the final state.


\subsection{Evolution of the global polarization}

Using the temperature and longitudinal velocity profiles provided by the hydrodynamic simulation, we are able to calculate the evolution of the quark polarization at a given position via Eq.~(\ref{eq:evoEq}).

\begin{figure}[tbp]
	\centering
	\begin{minipage}{0.49\linewidth}
		\centering
		\includegraphics[width=0.99\linewidth]{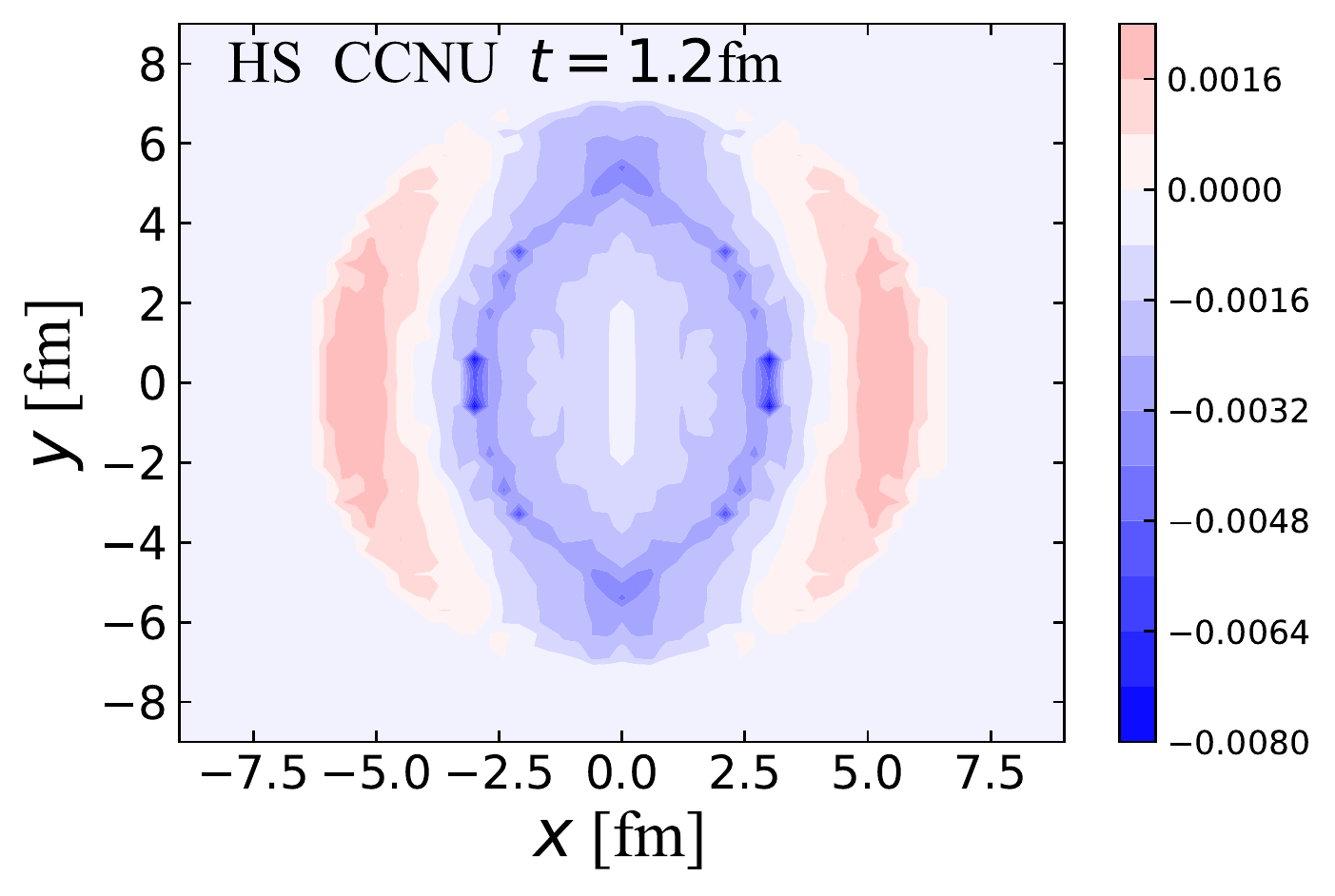}
	\end{minipage}
	\begin{minipage}{0.49\linewidth}
		\centering
		\includegraphics[width=0.99\linewidth]{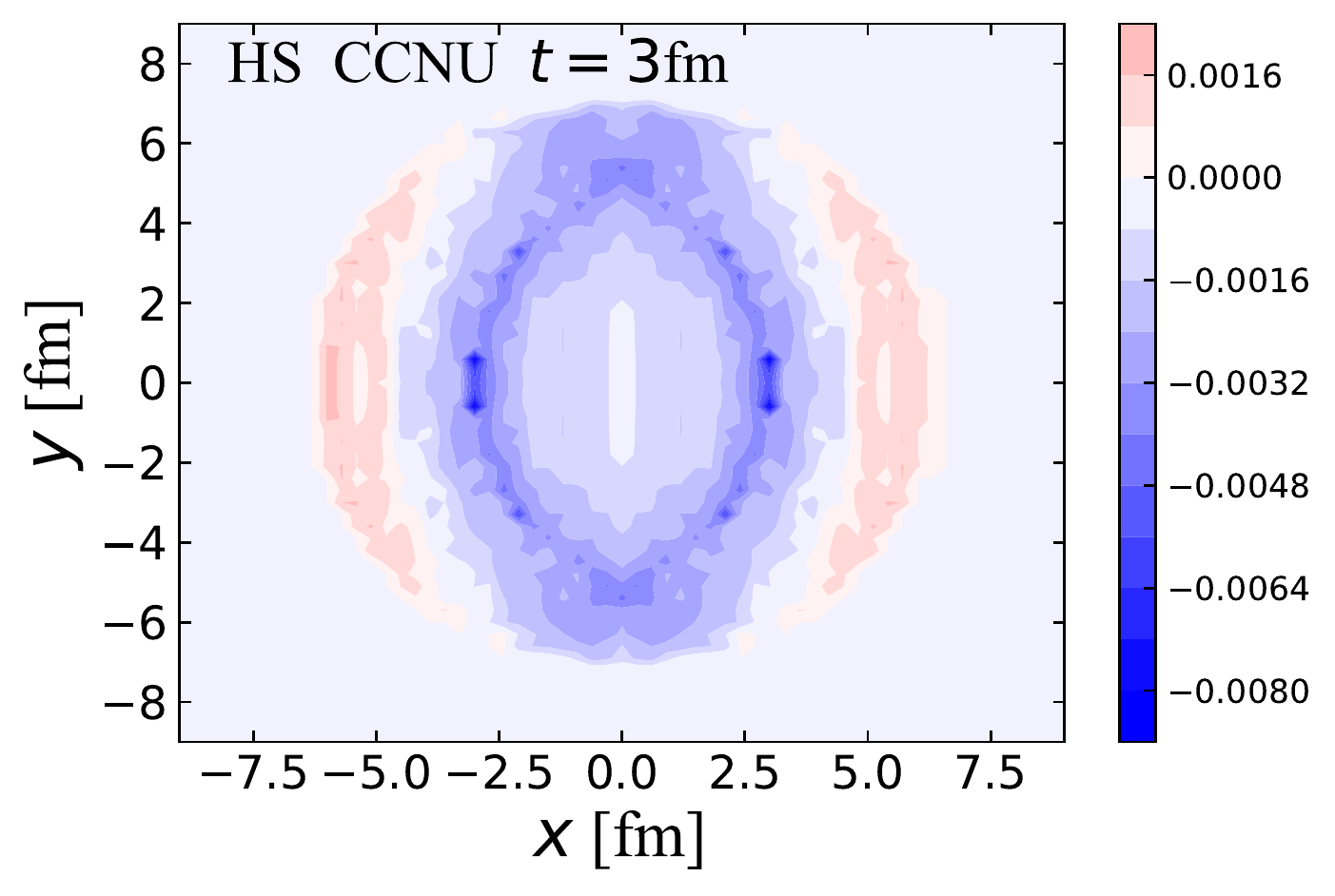}
	\end{minipage}
    \begin{minipage}{0.49\linewidth}
    	\centering
    	\includegraphics[width=0.99\linewidth]{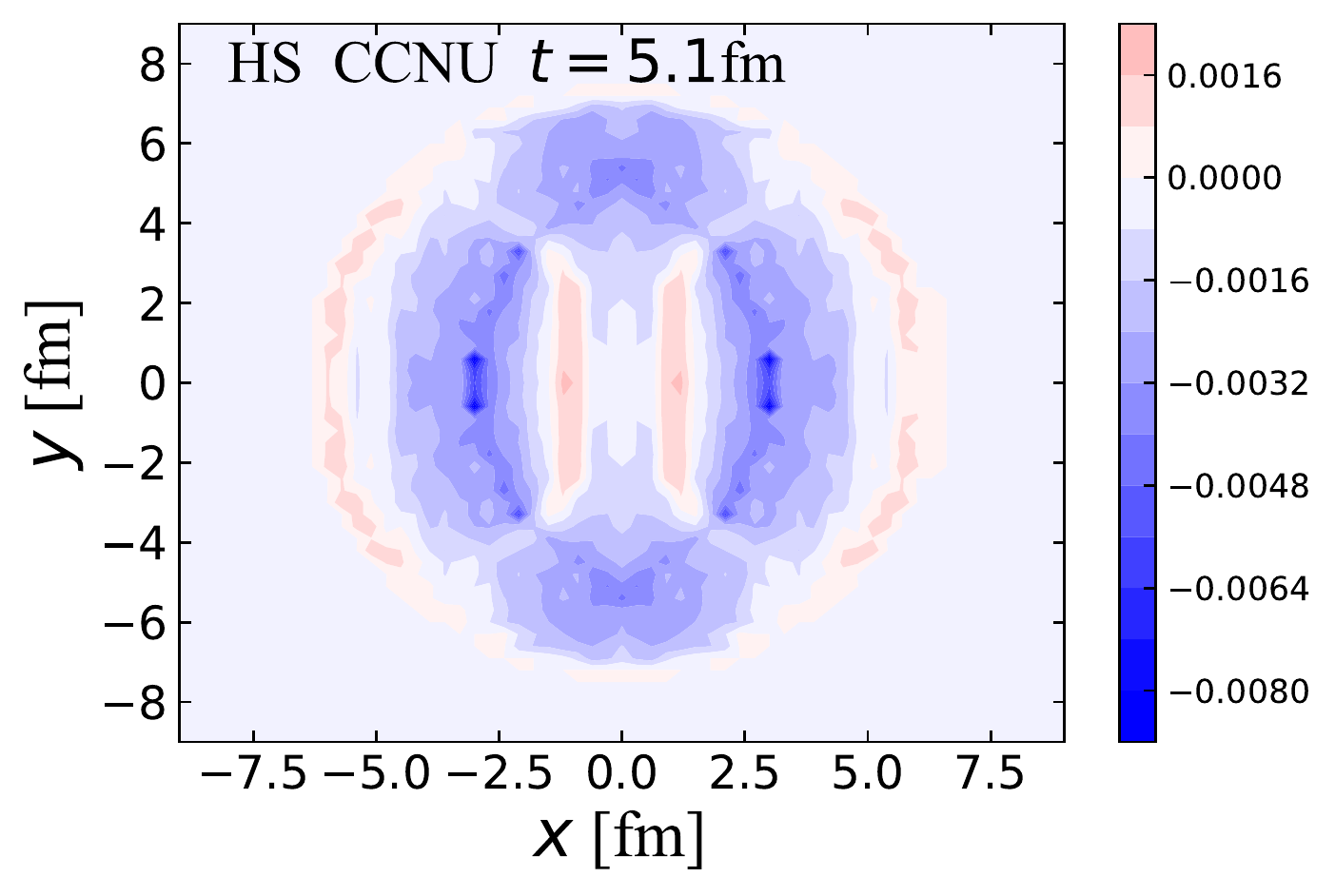}
    \end{minipage}
    \begin{minipage}{0.49\linewidth}
    	\centering
    	\includegraphics[width=0.99\linewidth]{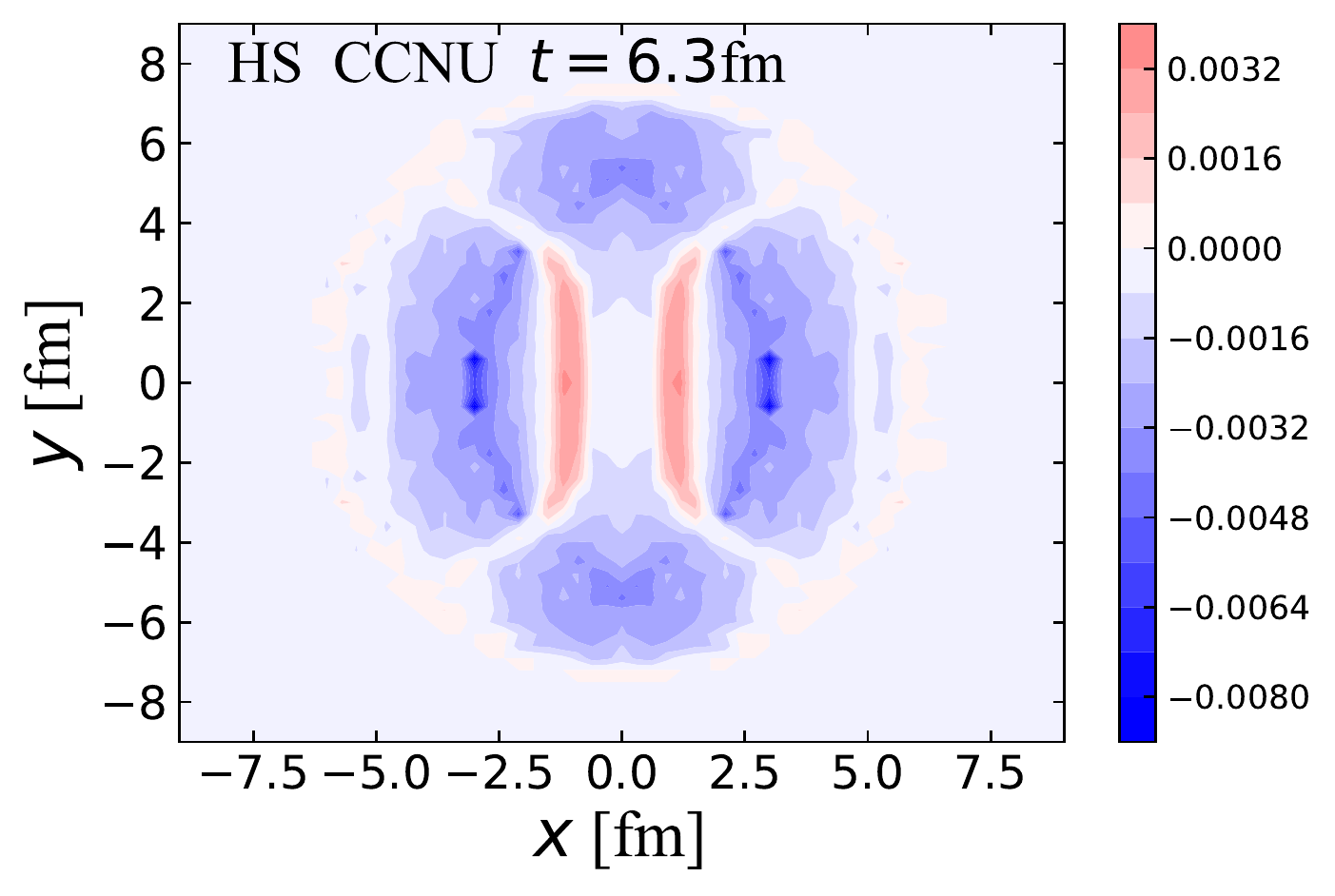}
    \end{minipage}
	
	\begin{minipage}{0.49\linewidth}
		\centering
		\includegraphics[width=0.99\linewidth]{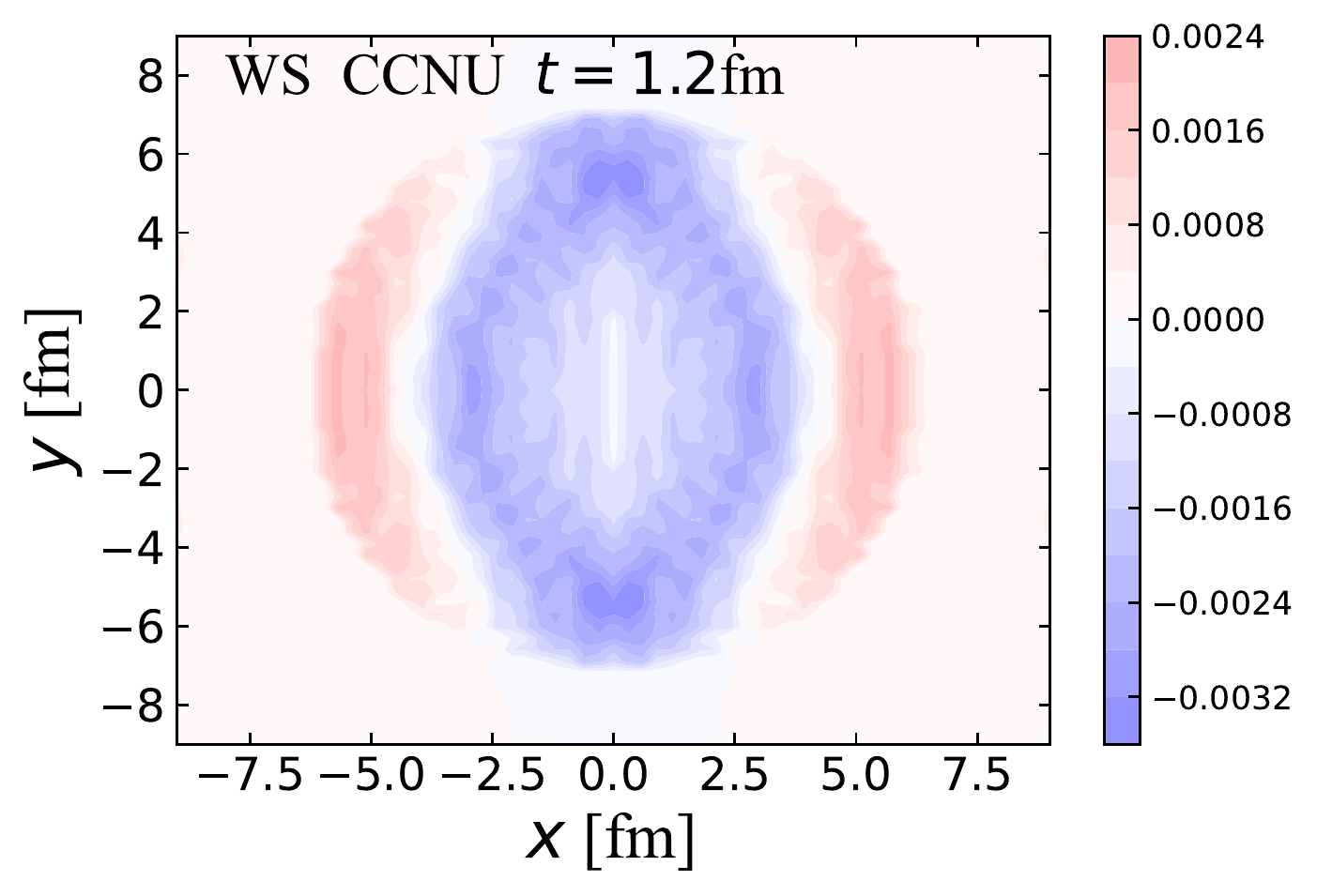}
	\end{minipage}
	\begin{minipage}{0.49\linewidth}
		\centering
		\includegraphics[width=0.99\linewidth]{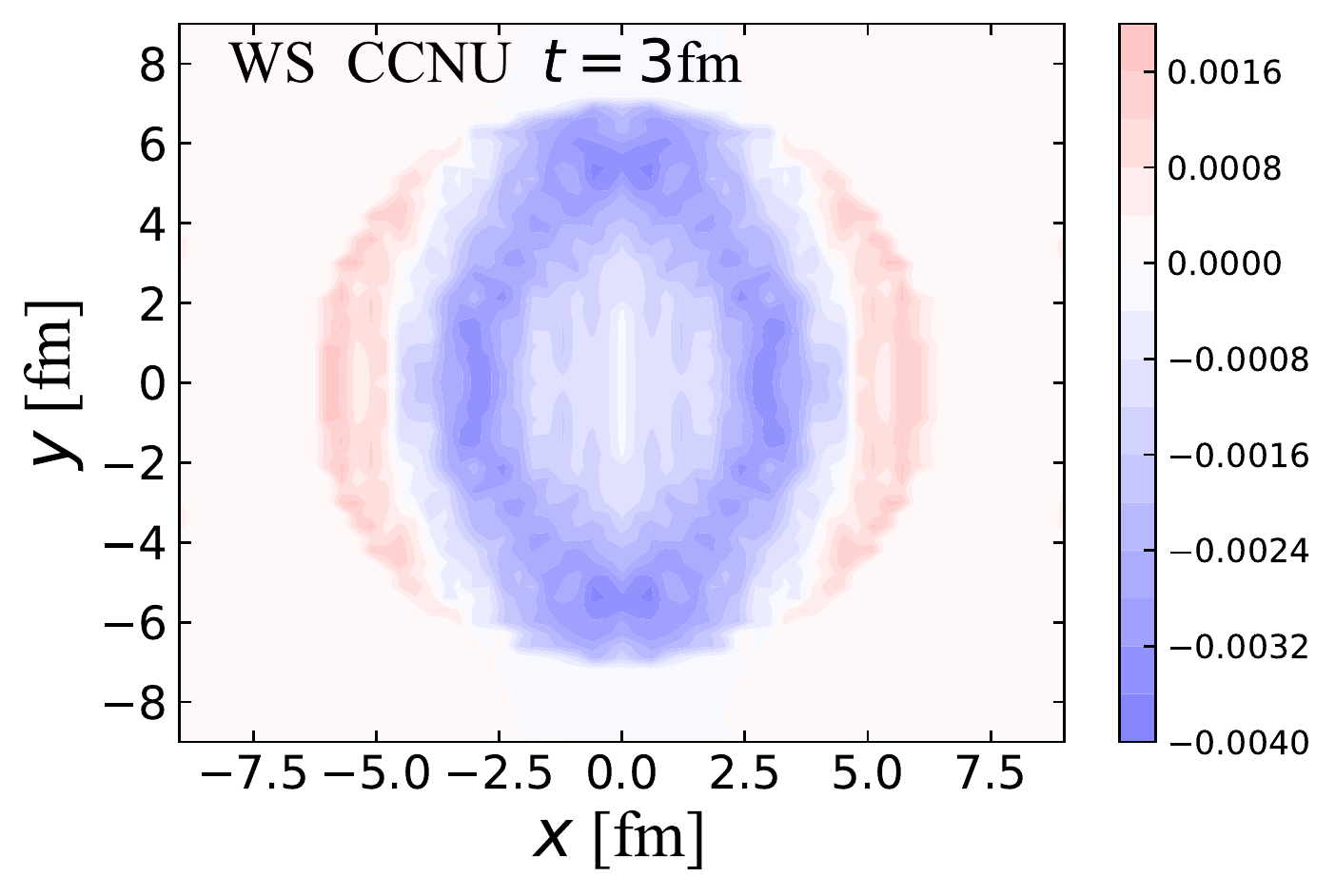}
	\end{minipage}
     \begin{minipage}{0.49\linewidth}
     	\centering
     	\includegraphics[width=0.99\linewidth]{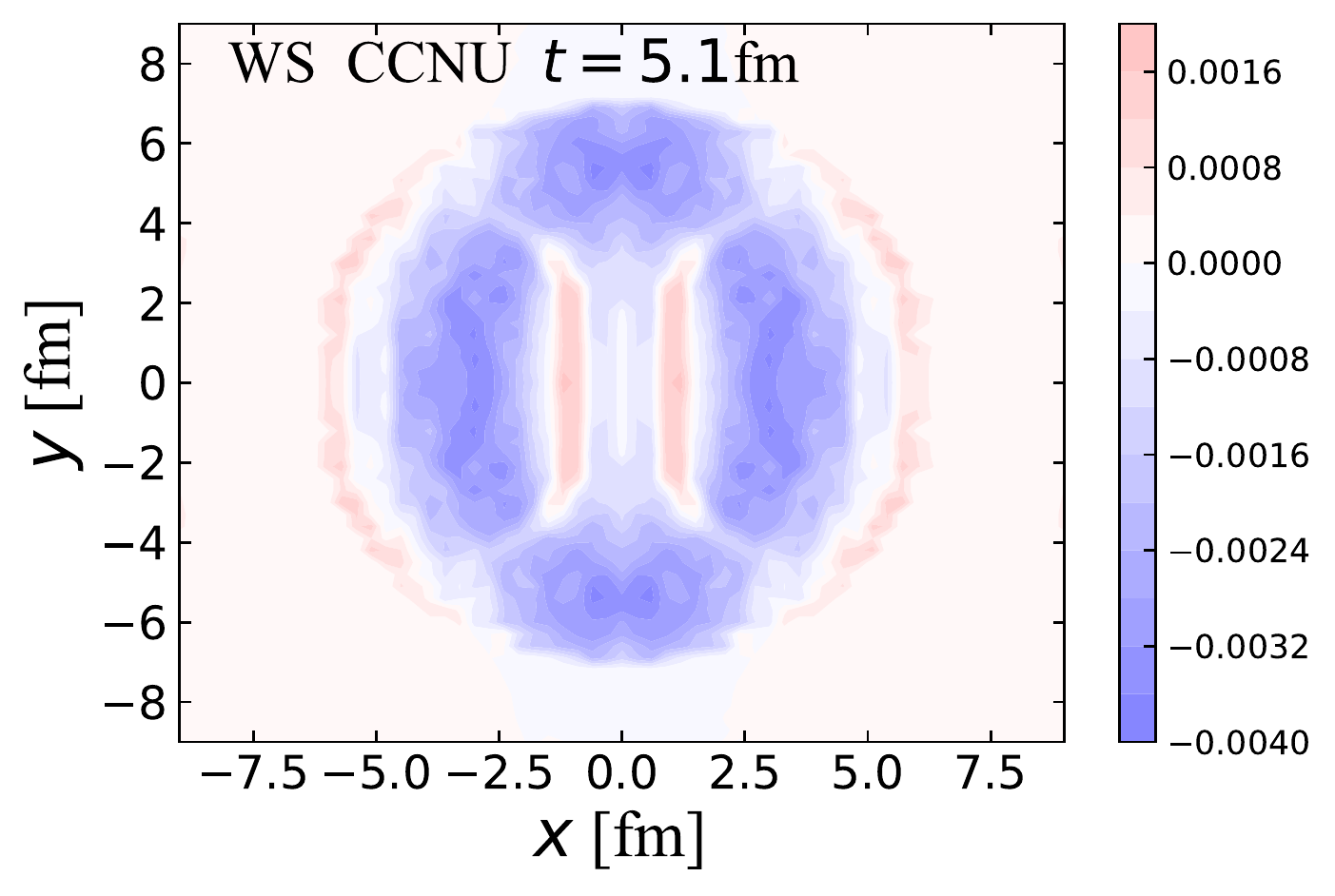}
     \end{minipage}
     \begin{minipage}{0.49\linewidth}
     	\centering
     	\includegraphics[width=0.99\linewidth]{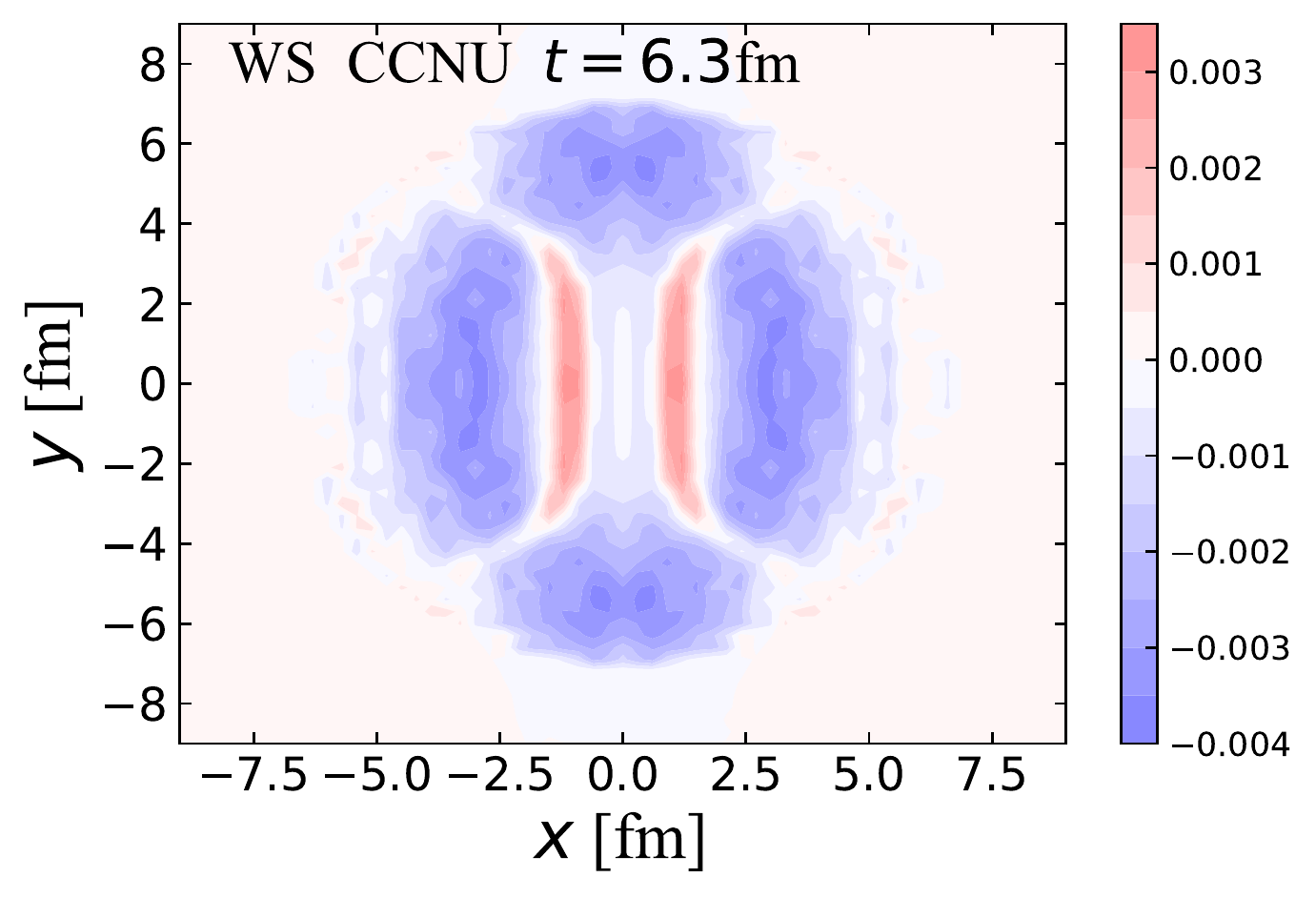}
     \end{minipage}
\renewcommand{\figurename}{FIG}
\caption{(Color online) Spatial distribution of quark polarization in the transverse plane at different times, compared between Hard Sphere (HS) and Woods-Saxon (WS) distributions of the initial nucleon density. The hydrodynamic evolution is initialized with the CCNU model.}
\label{fig:ccnuP}
\end{figure}

\begin{figure}[tbp]
	\centering
	\begin{minipage}{0.49\linewidth}
		\centering
		\includegraphics[width=0.99\linewidth]{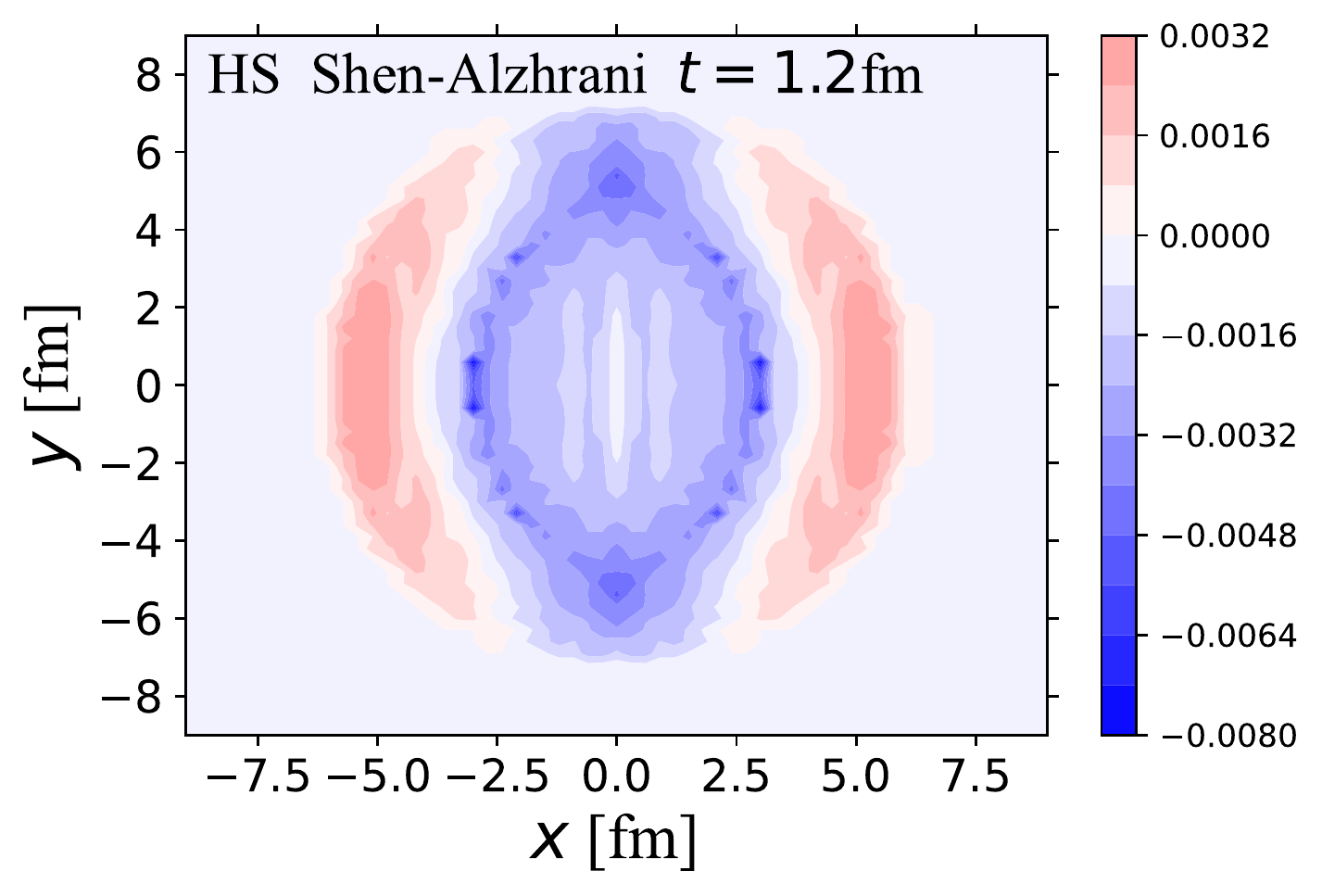}
	\end{minipage}
	\begin{minipage}{0.49\linewidth}
		\centering
		\includegraphics[width=0.99\linewidth]{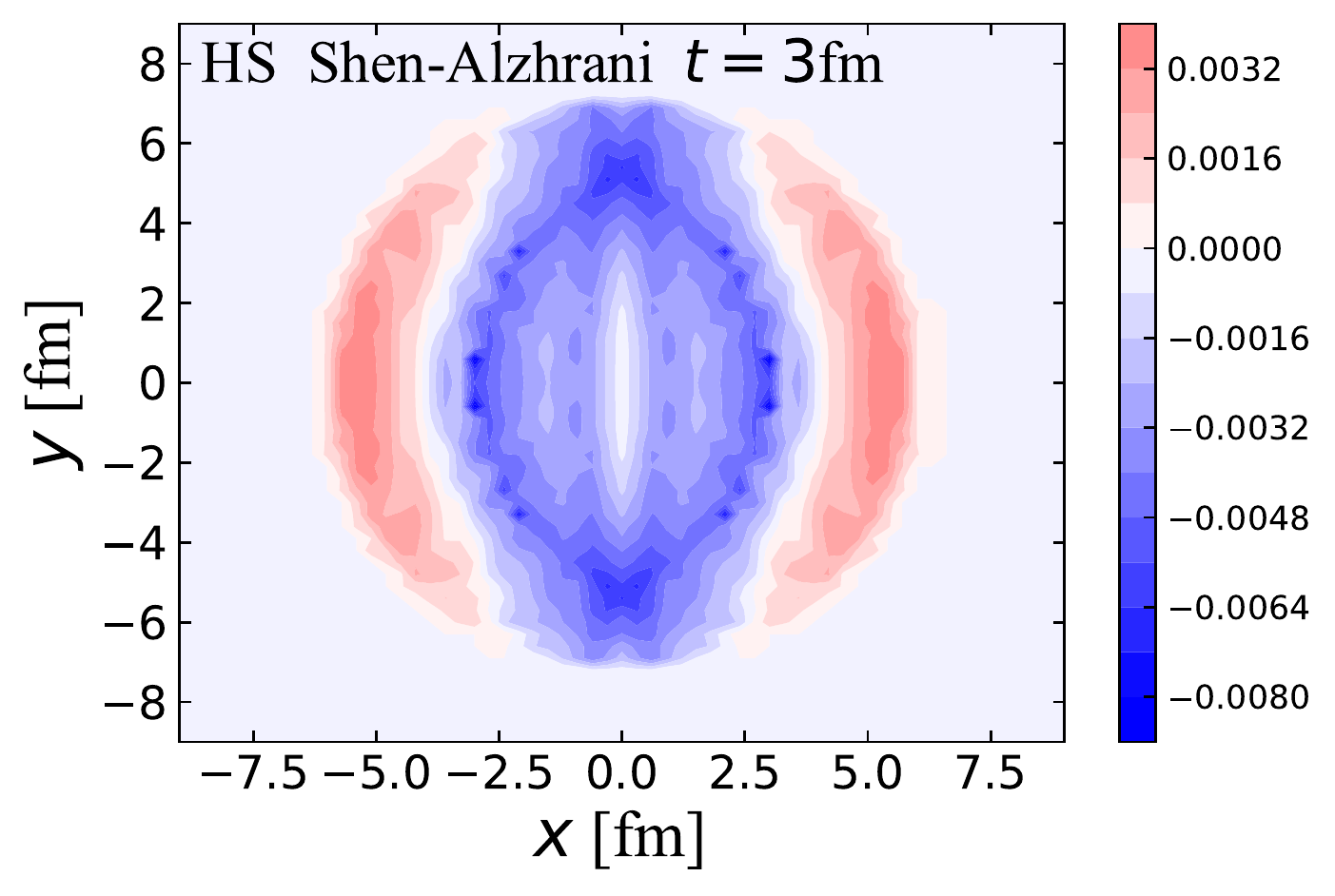}
	\end{minipage}
	\begin{minipage}{0.49\linewidth}
		\centering
		\includegraphics[width=0.99\linewidth]{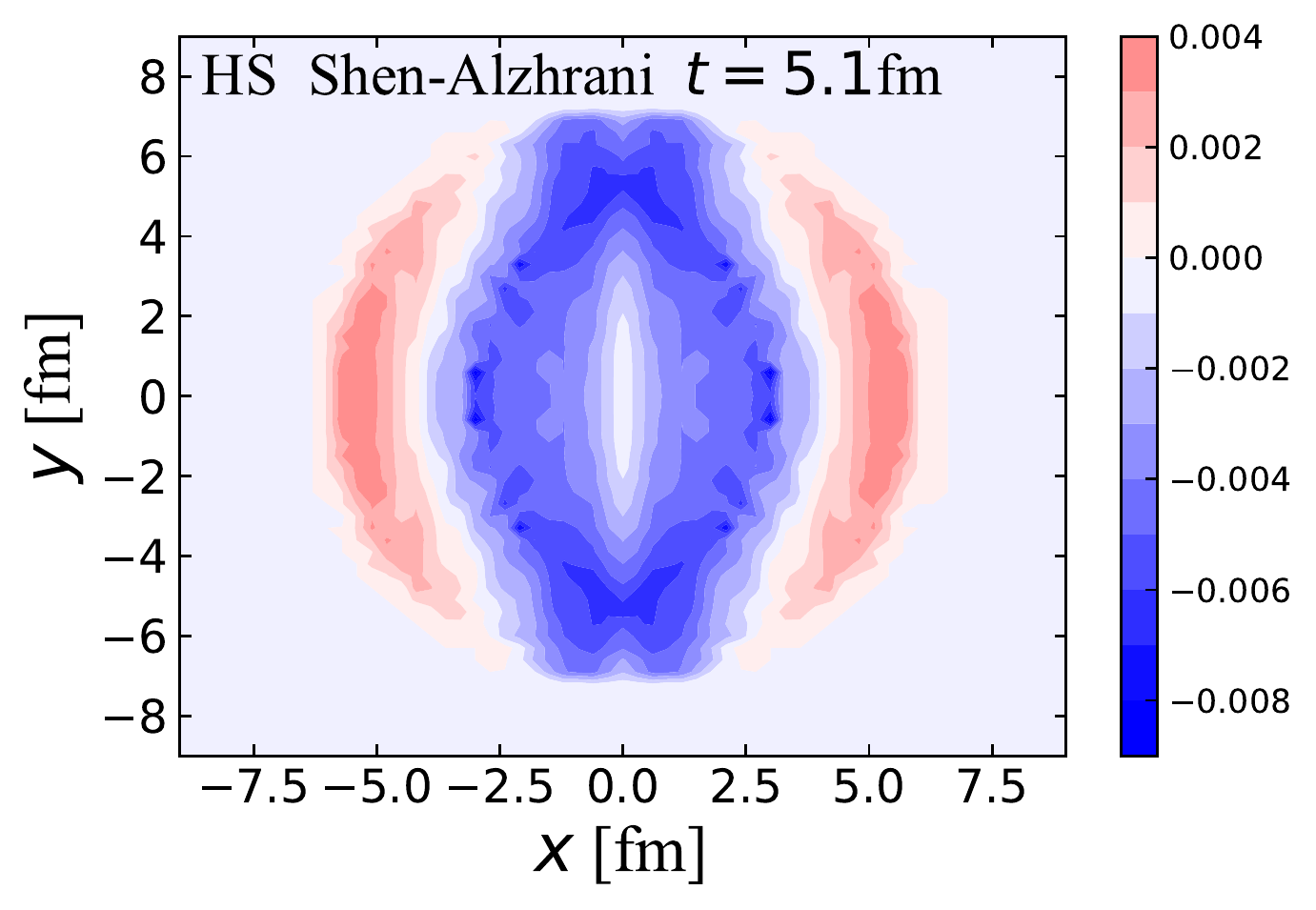}
	\end{minipage}
    \begin{minipage}{0.49\linewidth}
    	\centering
    	\includegraphics[width=0.99\linewidth]{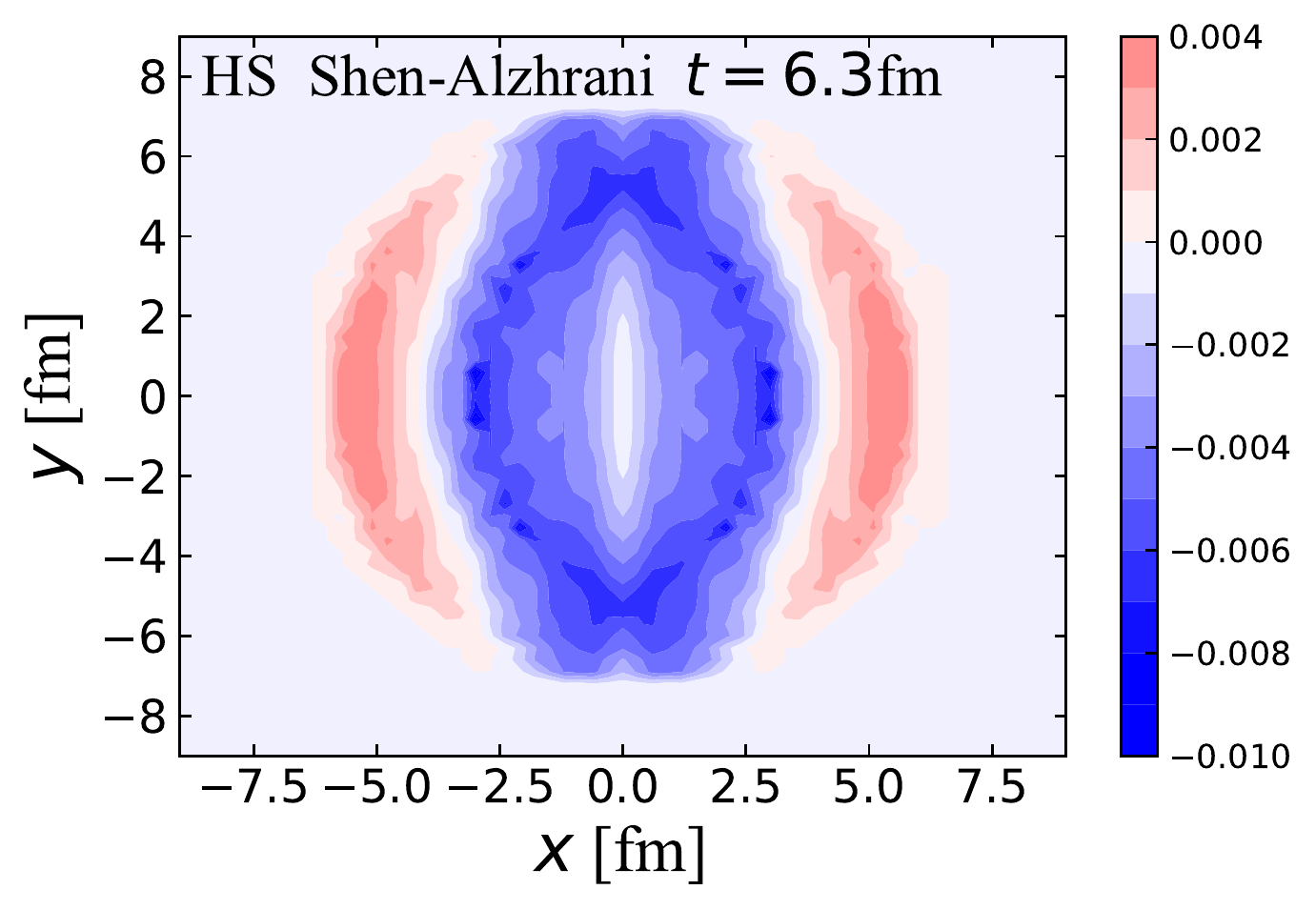}
    \end{minipage}

	\begin{minipage}{0.49\linewidth}
		\centering
		\includegraphics[width=0.99\linewidth]{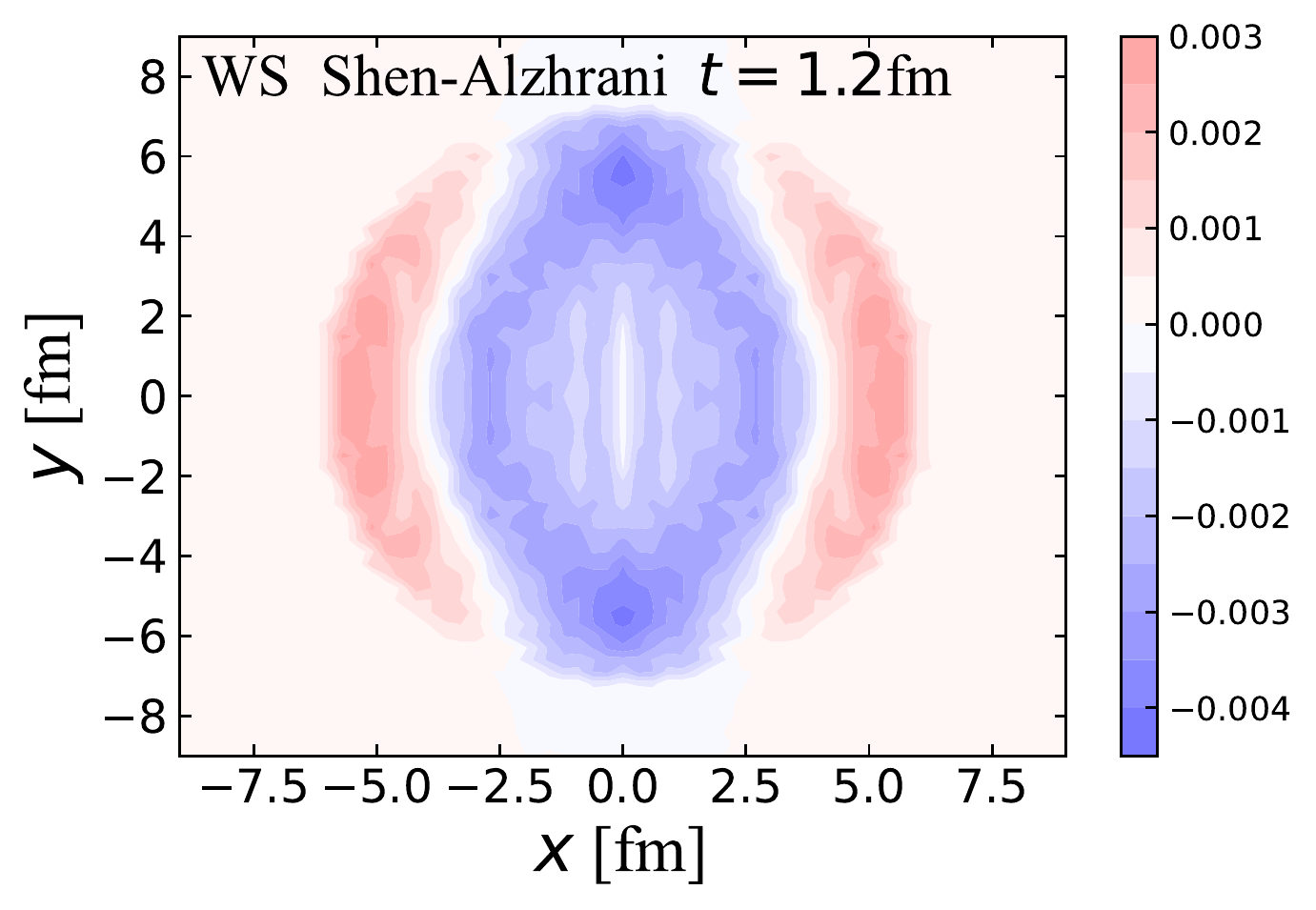}
	\end{minipage}
	\begin{minipage}{0.49\linewidth}
		\centering
		\includegraphics[width=0.99\linewidth]{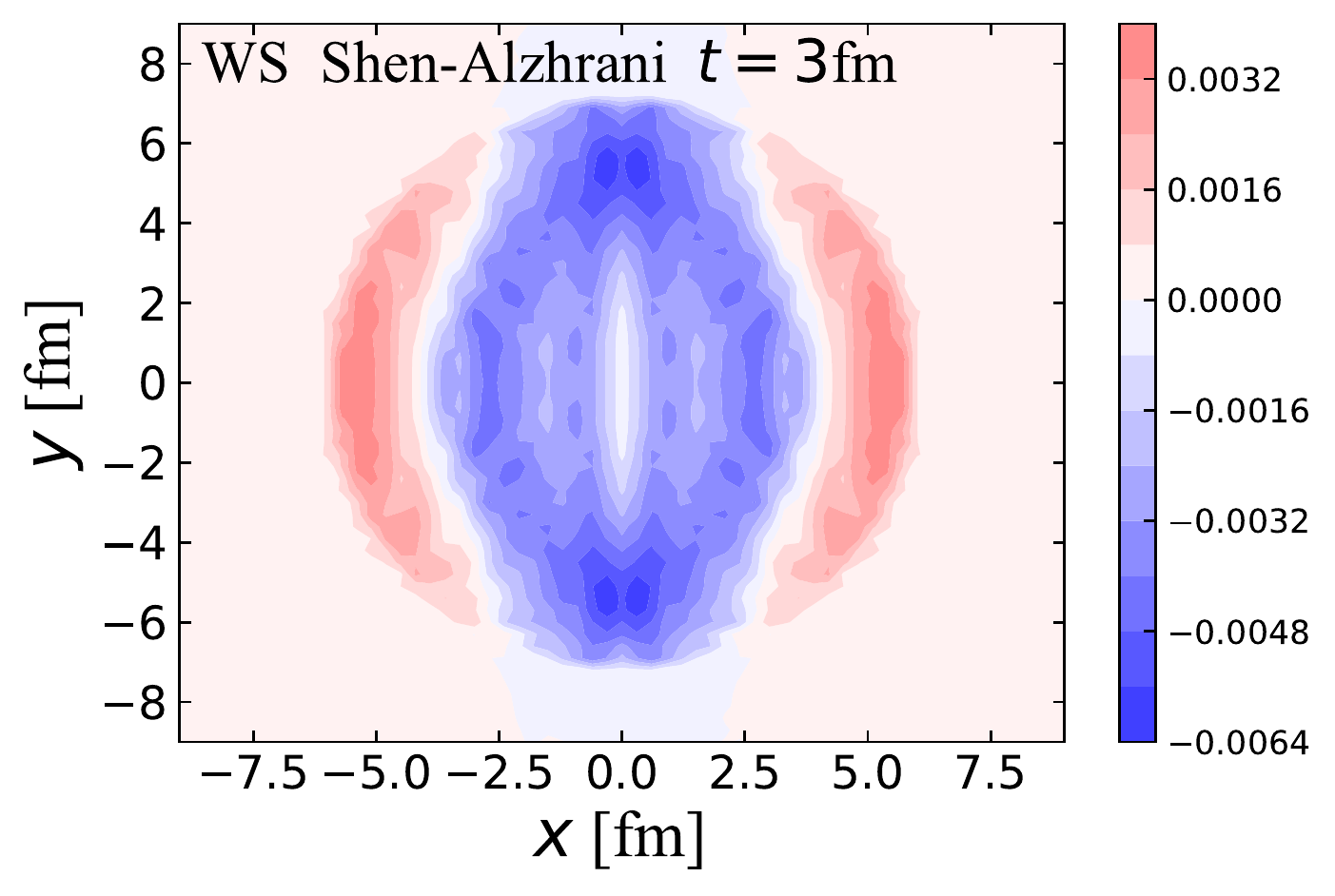}
	\end{minipage}
    \begin{minipage}{0.49\linewidth}
    	\centering
    	\includegraphics[width=0.99\linewidth]{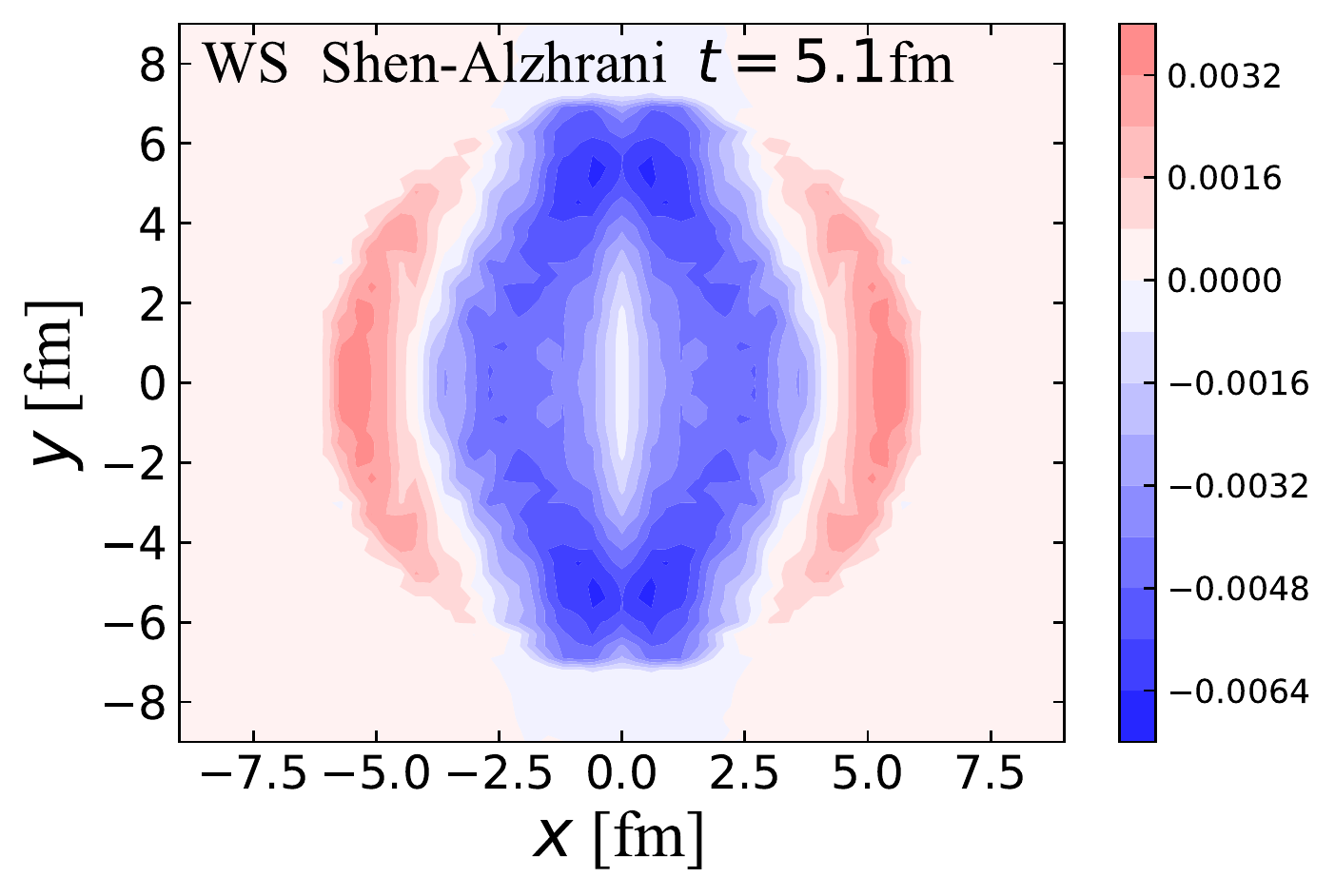}
    \end{minipage}
    \begin{minipage}{0.49\linewidth}
    	\centering
    	\includegraphics[width=0.99\linewidth]{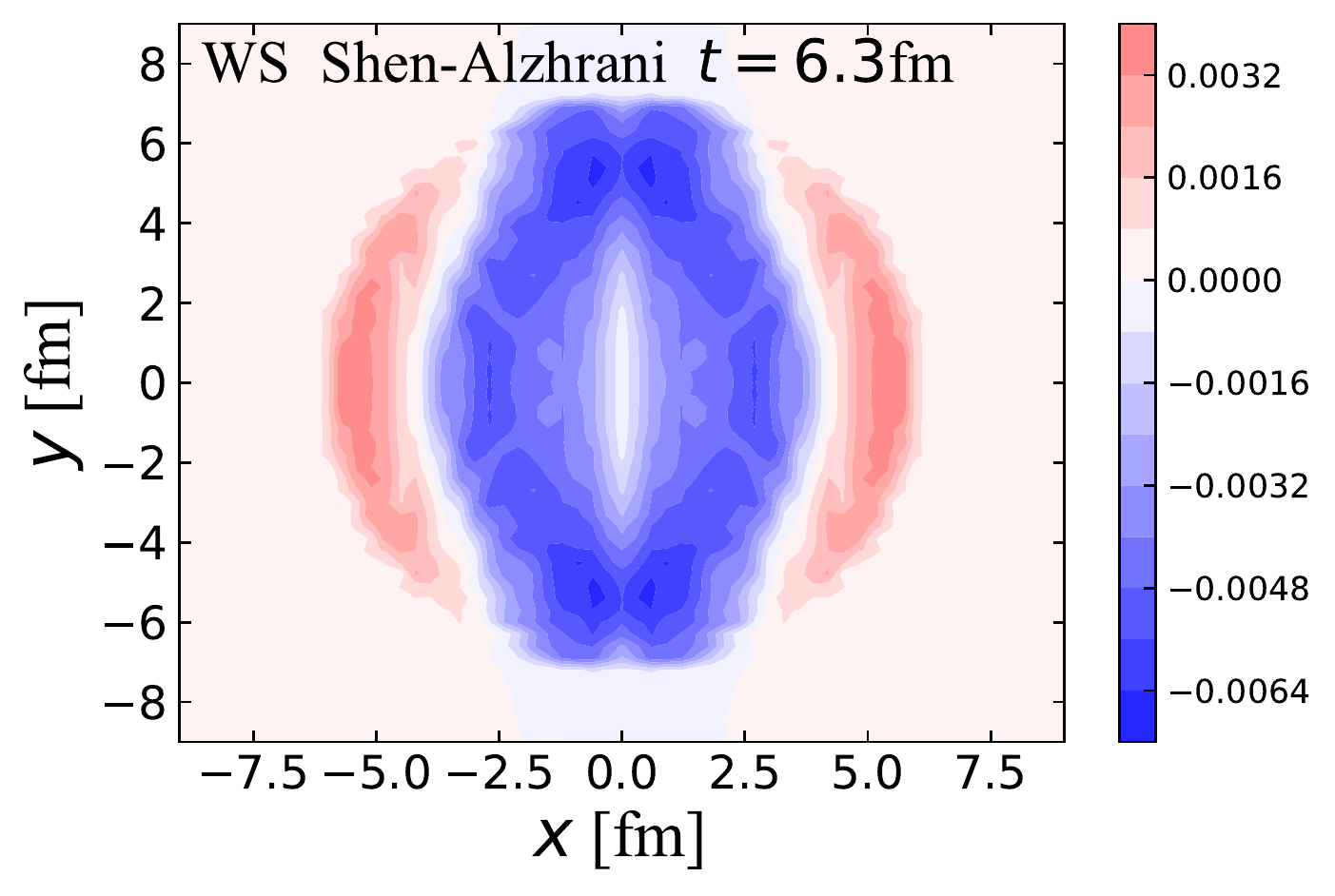}
    \end{minipage}
	\renewcommand{\figurename}{FIG}
	\caption{(Color online) Spatial distribution of quark polarization in the transverse plane at different times, compared between Hard Sphere (HS) and Woods-Saxon (WS) distributions of the initial nucleon density. The hydrodynamic evolution is initialized with the Shen-Alzhrani model.}
	\label{fig:shenP}
\end{figure}

%

Shown in Fig.~\ref{fig:ccnuP} are the snapshots of the spatial distribution of the quark polarization at different times. The strong coupling constant is taken as $g=2$ ($\alpha_\mathrm{s}=0.3$) for these calculations. The upper four panels evolve from the initial polarization evaluated with the Hard Sphere nucleon density distribution (the upper panel of Fig.~\ref{fb}), while the lower four panels are from the Woods-Saxon distribution (the lower panel of Fig.~\ref{fb}). The hydrodynamic expansion is initialized with the CCNU model. From the figure, one can see that as time evolves, the quark polarization generally increases in magnitude (or becomes more negative). Compared between the two nucleon density functions, we find a smoother distribution for the quark polarization across the transverse plane from the Woods-Saxon than the Hard Sphere distribution. It is interesting to note that positive values of polarization exist at locations far away from the QGP center not only for the Woods-Saxon distribution, but also for the Hard Sphere distribution. This is driven by the negative $\partial v_z/\partial x$ gradient at large $|x|$ at the initial time (as shown in Fig.~\ref{Vzx}). As time evolves, these positive values can disappear at locations far away from the QGP center, but start to appear near the center, which can be understood with the sign flip of $\partial v_z/\partial x$ both at large $|x|$ and around $x=0$ during the hydrodynamic expansion that starts with the CCNU initial condition (middle panel of Fig.~\ref{Vzx}).
At each time step, the polarization of each quark is only affected by fluid cells with local temperature above the freeze-out temperature $T_\mathrm{frz}$ here. We have verified that no visible difference can be observed between the CCNU and the Boz$\dot{\textrm{e}}$k-Wyskiel initial condition. In Fig.~\ref{fig:shenP}, we present the similar snapshots of quark polarization to Fig.~\ref{fig:ccnuP}, except that the hydrodynamic simulation starts with the initial energy density from the Shen-Alzhrani model. Compared between Fig.~\ref{fig:shenP} and Fig.~\ref{fig:ccnuP}, we observe a quicker increase (in magnitude) of the quark polarization within the QGP regime, which could be understood with the slower decay of the longitudinal flow velocity gradient with the Shen-Alzhrani initialization than the CCNU initialization, as we previously discussed in Fig.~\ref{Vzx}. No inversion is observed for either the positive value of polarization away from the center or the negative value near the center here, because the sign of $\partial v_z/\partial x$ remains during the hydrodynamic evolution if it is initialized with the Shen-Alzhrani model.

%

In the end, we calculate the average polarization over the entire transverse plane. We assume that when the local temperature drops below $T_\mathrm{frz}$, the fluid cell hadronizes and stops participating in subsequent scatterings. Therefore, the corresponding polarization also freezes. For each time step of the hydrodynamic evolution, we first calculate the change of polarization at each location as $\Delta P(x,y,t)$. The change of the average polarization within this time step is then contributed by fluid cells above $T_\mathrm{frz}$ and is given by
\begin{align}
	\overline{\Delta P}(t) = \frac{ \int_{\,T>T_\mathrm{frz}} dxdy s(x,y,t)\Delta P(x,y,t) }{ \int dxdy s(x,y,t)}.
\end{align}
This $\overline{\Delta P}(t)$ is then applied to calculate the time evolution of the transverse-plane-averaged polarization as
\begin{equation}
	\overline{P}(t+\Delta t) = \overline{P}(t) + \overline{\Delta P}(t),
\end{equation}
which starts from an initial value produced by the primordial hard scatterings (shown in Fig.~\ref{fig:initialP}).

\begin{figure}[t!]
	\begin{center}
		\includegraphics[width=0.4\textwidth]{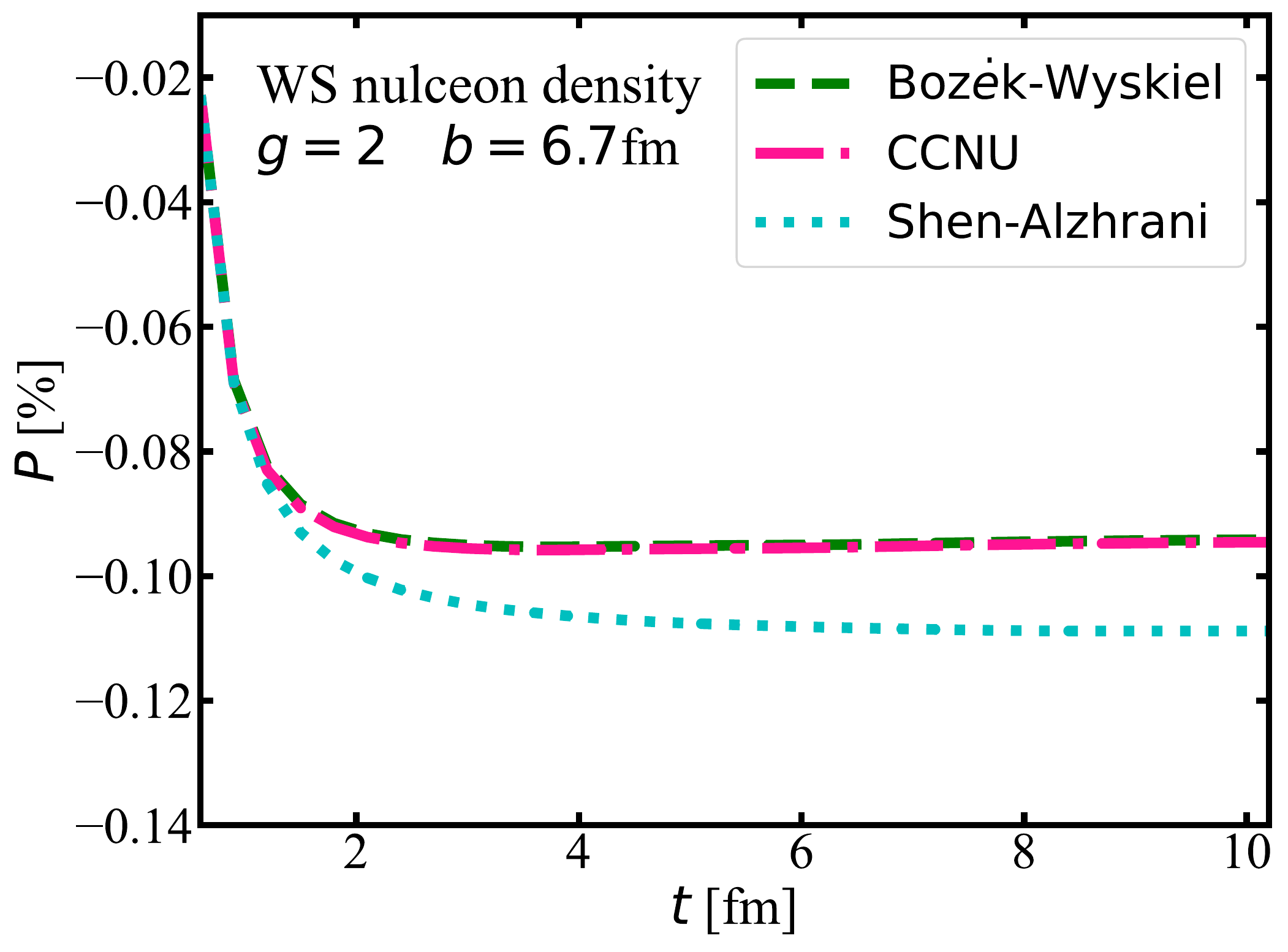}		
		\includegraphics[width=0.4\textwidth]{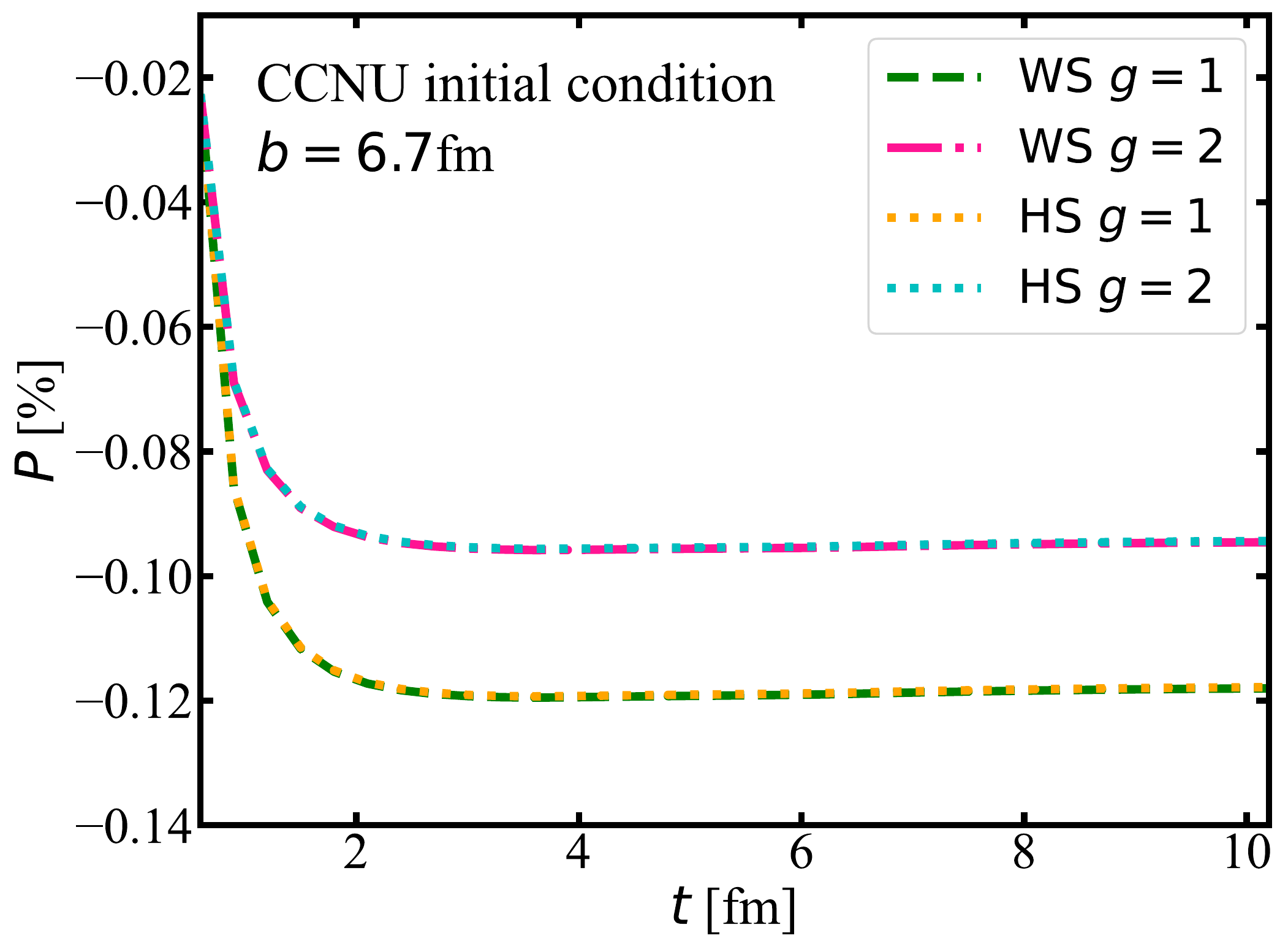}
		\includegraphics[width=0.4\textwidth]{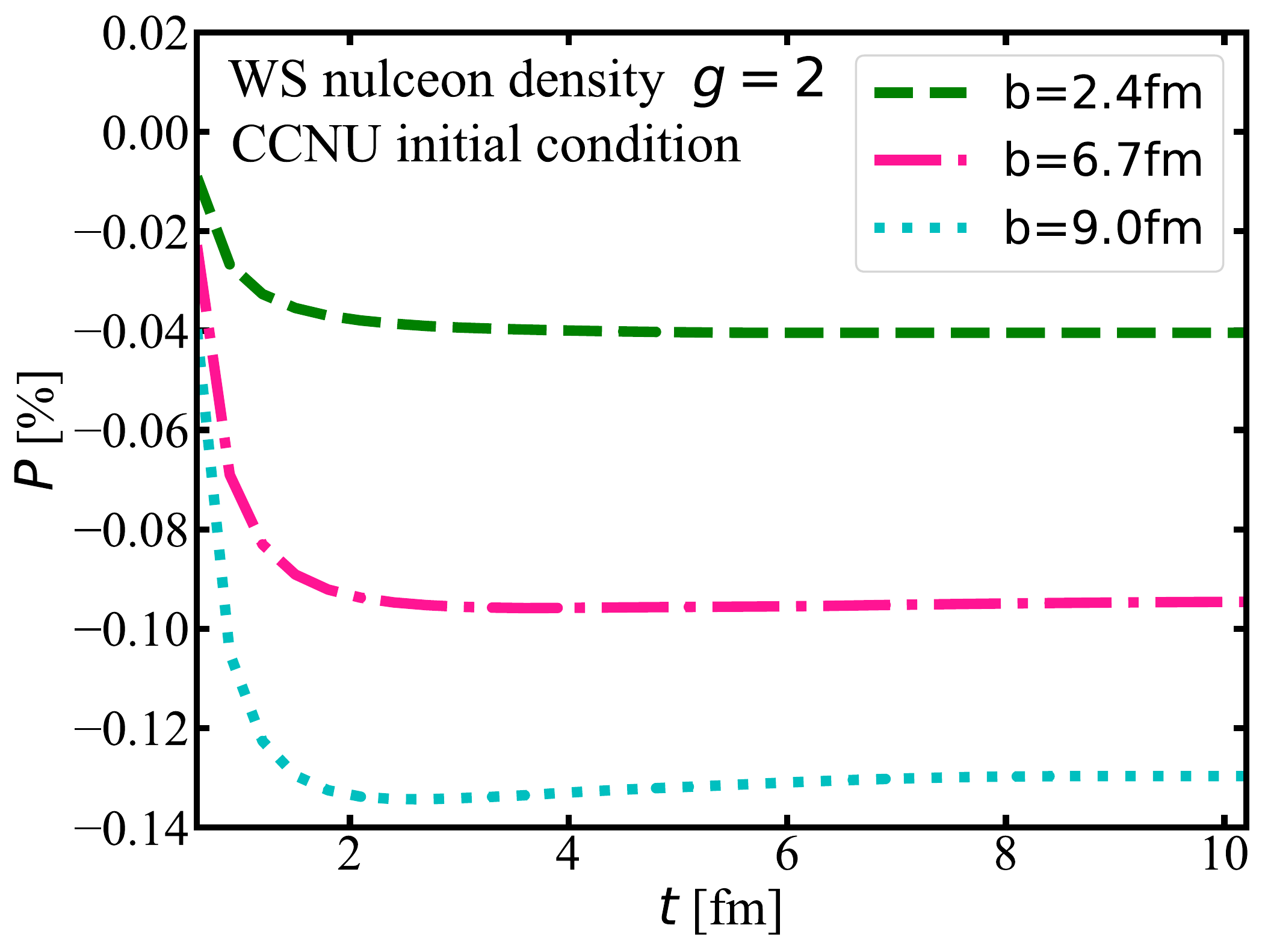}	
	\end{center}
	\caption{(Color online) Time evolution of the transverse-plane-averaged polarization, compared between using different initial conditions for the hydrodynamic evolution (upper panel), different density functions of the initial nucleon distribution and different values of $g$ (middle panel), and different impact parameters (lower panel).}
	\label{pf}
\end{figure}

In Fig.~\ref{pf} we present the time evolution of the transverse-plane-averaged polarization at mid-spacetime-rapidity. Although positive values of polarization can exist at certain spacetime, due to the relatively small entropy (energy) densities at these locations, the average polarization is still negative as expected. In the upper panel, we use $g=2$ as for Figs.~\ref{fig:ccnuP} and~\ref{fig:shenP}, and compare between hydrodynamic expansion with different initial conditions. One observes that the fluid velocity profiles of the QGP have a non-negligible impact on the final average polarization. The tilted geometry of the initial energy density, as implemented in the Boz$\dot{\textrm{e}}$k-Wyskiel and CCNU models, leads to a slower increase (in magnitude) of polarization compared to that from the Shen-Alzhrani initialization. This can be understood with the faster decay of the longitudinal velocity gradient in the former case than the latter, as discussed earlier. The dependence of the global polarization on the initial condition of the QGP has also been discussed in Refs.~\cite{Becattini:2015ska,STAR:2018gyt}. We also notice that the global polarization mainly develops during the first 2~fm of the QGP evolution due to the large $v_z$ gradient and energy density at the early time. The global polarization we obtain for the final state is around 0.095\%$\sim$0.109\% (about 15\% uncertainty), depending on the selected initial condition. This is qualitatively consistent with (though quantitatively smaller than) the $\Lambda$ polarization -- $0.277 \pm 0.040 ~\mathrm{(stat)} \pm\; ^{0.039}_{0.049}~\mathrm{(sys)}$~[\%] -- observed at the top RHIC energy~\cite{STAR:2018gyt}. The magnitude of polarization we obtain here using a (3+1)-D relativistic hydrodynamic model is much smaller than that in the earlier study~\cite{Huang:2011ru} using the relativistic laminar flow model, because of the weaker longitudinal flow velocity from our simulation.

In the middle panel of Fig.~\ref{pf}, we investigate effects of the initial nucleon density distribution and the strong coupling constant on the final-state global polarization. Although difference between the Hard Sphere and Woods-Saxon distributions could be observed in the spatial distribution of polarization previously in Figs.~\ref{fig:ccnuP} and~\ref{fig:shenP}, it can be hardly seen here after the polarization has been averaged over the transverse plane. As discussed in Sec.~\ref{sec2}, the strong coupling constant can affect the quark polarization through the Debye screening mass. Since $\mu$ appears on both the numerator and the denominator of Eq.~(\ref{eq:deltaP}), and the value of polarization can be either positive or negative across the transverse plane, the $g$-dependence of the quark polarization differs from location to location. After averaging over the entire transverse plane, we find a larger magnitude of the quark polarization when $g$ is reduced from 2 to 1.

In the lower panel of Fig.~\ref{pf}, we study the impact parameter dependence of this time evolution of polarization. As the impact parameter increases, a significantly larger magnitude of global polarization is obtained. This is consistent with the stronger orbital angular momentum deposited into the nuclear matter in more peripheral collisions. Note that as $b$ increases, not only the initial polarization from the primordial hard scatterings becomes larger, as was also shown in Fig.~\ref{fig:initialP}, the increase of polarization during the QGP evolution becomes stronger as well because of the larger fluid velocity gradient $\partial v_z/\partial x$ formed in more peripheral collisions.

\section{Summary and outlook}
\label{sec4}

We have investigated the production and evolution of the global polarization of quarks in relativistic heavy-ion collisions within a perturbative approach. The spin-independent and dependent parts of the quark-potential scattering cross sections have been consistently applied to both the initial hard scatterings between colliding nuclei and the subsequent quark scatterings through the QGP. Compared to earlier studies, we have improved this perturbative approach in two aspects: (1) the two-body scattering model where the projectile quark is constrained in a half hemisphere relative to the target potential has been extended to realistic spatial distributions of targets and projectiles; and (2) a (3+1)-D viscous hydrodynamic model is adopted for simulating the QGP expansion. Effects of the nucleon density function and the initial geometry of the QGP medium on the final-state quark polarization have been explored in detail.

Within this improved approach, we have found that the spatial distribution of quark polarization depends on the nucleon density function inside the colliding nuclei. While polarization sharply centers around the edge of the overlapping region between the colliding nuclei with the Hard Sphere model, a smoother distribution across the transverse plane with possible positive values far away from the overlapping region can be seen with the Woods-Saxon model. However, after averaging over the transverse plane, these two density functions provide consistent magnitudes of the quark polarization, except for very peripheral collisions. Three different setups of the initial energy density distributions of the QGP have been compared in this work. It has been found that with a counter-clockwise tilted initial geometry in the reaction plane, the QGP expansion leads to opposite longitudinal flow velocity ($v_z$) to its initial direction. This can accelerate the decay of $v_z$ and even reverse its direction at late time, thus resulting in a smaller magnitude of quark polarization compared to calculation without using the tilted initial condition. The transverse-plane-averaged global polarization we obtain is about 0.095\%$\sim$0.109\% at mid-spacetime-rapidity when the strong coupling constant is taken as $g=2$ inside the QGP. The approximately 15\% uncertainty quantifies the sensitivity of the global polarization to the longitudinal flow velocity profiles of the QGP, and may serve as a novel tool to help constrain the initial energy density distribution of the QGP in the future, when both theoretical calculations and experimental measurements become more precise. 

While our study constitutes a step forward in a more quantitative understanding of the production and evolution of the global polarization using perturbative calculations, it should be further extended in several directions. For instance, our current calculation is limited at the quark level. A sophisticated hadronization scheme~\cite{Sheng:2020ghv,Sun:2017xhx} and decay contributions to polarized resonant states~\cite{Karpenko:2016jyx} should be introduced to connect our current result to the realistic polarization of $\Lambda$ hyperons measured by experiments. In addition, it is also important to study the beam energy and rapidity dependences of the global polarization, as have already been measured by the STAR experiments~\cite{STAR:2017ckg,STAR:2018gyt}. To achieve this, one may also need to release the small angle approximation applied in our current calculation [Eqs.~(\ref{eq:I0simplify}) and~(\ref{eq:I1})], which might not be valid when the center-of-mass energy of a colliding quark pair is small~\cite{Gao:2007bc}. We will address these aspects in our follow-up efforts.


\begin{acknowledgements}
We are grateful for helpful discussions with Zuo-Tang Liang, Xu-Guang Huang and Xiang-Yu Wu. This work was supported by the National Natural Science Foundation of China (NSFC) under Grant Nos.~12175122, 2021-867 and 11935007, the  Natural Science Foundation of Shandong Province under Grant No.~ZR2020MA099, Guangdong Major Project of Basic and Applied Basic Research No.~2020B0301030008, the Natural Science Foundation of Hubei Province No.~2021CFB272, the Education Department of Hubei Province of China with Young Talents Project No.~Q20212703, the open foundation of Key Laboratory of Quark and Lepton Physics (MOE) No.~QLPL2021P01 and
the Xiaogan Natural Science Foundation under Grant No.~XGKJ2021010016. 
\end{acknowledgements}

\bibliographystyle{unsrt}
\bibliography{SCrefs}

\begin{thebibliography}{10}

\bibitem{Gyulassy:2004zy}
M.~Gyulassy and L.~McLerran.
\newblock {New forms of QCD matter discovered at RHIC}.
\newblock {\em Nucl. Phys. A}, 750:30--63, 2005.

\bibitem{Jacobs:2004qv}
Peter Jacobs and Xin-Nian Wang.
\newblock {Matter in extremis: Ultrarelativistic nuclear collisions at RHIC}.
\newblock {\em Prog. Part. Nucl. Phys.}, 54:443--534, 2005.

\bibitem{Gao:2007bc}
Jian-Hua Gao, Shou-Wan Chen, Wei-tian Deng, Zuo-Tang Liang, Qun Wang, and
  Xin-Nian Wang.
\newblock {Global quark polarization in non-central A+A collisions}.
\newblock {\em Phys. Rev. C}, 77:044902, 2008.

\bibitem{Deng:2016gyh}
Wei-Tian Deng and Xu-Guang Huang.
\newblock {Vorticity in Heavy-Ion Collisions}.
\newblock {\em Phys. Rev. C}, 93(6):064907, 2016.

\bibitem{Jiang:2016woz}
Yin Jiang, Zi-Wei Lin, and Jinfeng Liao.
\newblock {Rotating quark-gluon plasma in relativistic heavy ion collisions}.
\newblock {\em Phys. Rev. C}, 94(4):044910, 2016.
\newblock [Erratum: Phys.Rev.C 95, 049904 (2017)].

\bibitem{Liang:2004ph}
Zuo-Tang Liang and Xin-Nian Wang.
\newblock {Globally polarized quark-gluon plasma in non-central A+A
  collisions}.
\newblock {\em Phys. Rev. Lett.}, 94:102301, 2005.
\newblock [Erratum: Phys.Rev.Lett. 96, 039901 (2006)].

\bibitem{Liang:2004xn}
Zuo-Tang Liang and Xin-Nian Wang.
\newblock {Spin alignment of vector mesons in non-central A+A collisions}.
\newblock {\em Phys. Lett. B}, 629:20--26, 2005.

\bibitem{Ipp:2007ng}
Andreas Ipp, Antonino Di~Piazza, Jorg Evers, and Christoph~H. Keitel.
\newblock {Photon polarization as a probe for quark-gluon plasma dynamics}.
\newblock {\em Phys. Lett. B}, 666:315--319, 2008.

\bibitem{STAR:2007ccu}
B.~I. Abelev et~al.
\newblock {Global polarization measurement in Au+Au collisions}.
\newblock {\em Phys. Rev. C}, 76:024915, 2007.
\newblock [Erratum: Phys.Rev.C 95, 039906 (2017)].

\bibitem{STAR:2017ckg}
L.~Adamczyk et~al.
\newblock {Global $\Lambda$ hyperon polarization in nuclear collisions:
  evidence for the most vortical fluid}.
\newblock {\em Nature}, 548:62--65, 2017.

\bibitem{STAR:2018gyt}
Jaroslav Adam et~al.
\newblock {Global polarization of $\Lambda$ hyperons in Au+Au collisions at
  $\sqrt{s_{_{NN}}}$ = 200 GeV}.
\newblock {\em Phys. Rev. C}, 98:014910, 2018.

\bibitem{Becattini:2020ngo}
Francesco Becattini and Michael~A. Lisa.
\newblock {Polarization and Vorticity in the Quark\textendash{}Gluon Plasma}.
\newblock {\em Ann. Rev. Nucl. Part. Sci.}, 70:395--423, 2020.

\bibitem{Becattini:2007nd}
F.~Becattini and F.~Piccinini.
\newblock {The Ideal relativistic spinning gas: Polarization and spectra}.
\newblock {\em Annals Phys.}, 323:2452--2473, 2008.

\bibitem{Becattini:2007sr}
F.~Becattini, F.~Piccinini, and J.~Rizzo.
\newblock {Angular momentum conservation in heavy ion collisions at very high
  energy}.
\newblock {\em Phys. Rev. C}, 77:024906, 2008.

\bibitem{Becattini:2013fla}
F.~Becattini, V.~Chandra, L.~Del~Zanna, and E.~Grossi.
\newblock {Relativistic distribution function for particles with spin at local
  thermodynamical equilibrium}.
\newblock {\em Annals Phys.}, 338:32--49, 2013.

\bibitem{Li:2017slc}
Hui Li, Long-Gang Pang, Qun Wang, and Xiao-Liang Xia.
\newblock {Global $\Lambda$ polarization in heavy-ion collisions from a
  transport model}.
\newblock {\em Phys. Rev. C}, 96(5):054908, 2017.

\bibitem{Li:2021zwq}
Hui Li, Xiao-Liang Xia, Xu-Guang Huang, and Huan~Zhong Huang.
\newblock {Global spin polarization of multistrange hyperons and feed-down
  effect in heavy-ion collisions}.
\newblock {\em Phys. Lett. B}, 827:136971, 2022.

\bibitem{Huang:2020dtn}
Xu-Guang Huang, Jinfeng Liao, Qun Wang, and Xiao-Liang Xia.
\newblock {Vorticity and Spin Polarization in Heavy Ion Collisions: Transport
  Models}.
\newblock {\em Lect. Notes Phys.}, 987:281--308, 2021.

\bibitem{Betz:2007kg}
Barbara Betz, Miklos Gyulassy, and Giorgio Torrieri.
\newblock {Polarization probes of vorticity in heavy ion collisions}.
\newblock {\em Phys. Rev. C}, 76:044901, 2007.

\bibitem{Pang:2016igs}
Long-Gang Pang, Hannah Petersen, Qun Wang, and Xin-Nian Wang.
\newblock {Vortical Fluid and $\Lambda$ Spin Correlations in High-Energy
  Heavy-Ion Collisions}.
\newblock {\em Phys. Rev. Lett.}, 117(19):192301, 2016.

\bibitem{Fang:2016vpj}
Ren-hong Fang, Long-gang Pang, Qun Wang, and Xin-nian Wang.
\newblock {Polarization of massive fermions in a vortical fluid}.
\newblock {\em Phys. Rev. C}, 94(2):024904, 2016.

\bibitem{Karpenko:2016jyx}
I.~Karpenko and F.~Becattini.
\newblock {Study of $\Lambda $ polarization in relativistic nuclear collisions
  at $\sqrt{s_\mathrm {NN}}=7.7$ \textendash{}200 GeV}.
\newblock {\em Eur. Phys. J. C}, 77(4):213, 2017.

\bibitem{Baznat:2017jfj}
Mircea Baznat, Konstantin Gudima, Alexander Sorin, and Oleg Teryaev.
\newblock {Hyperon polarization in heavy-ion collisions and holographic
  gravitational anomaly}.
\newblock {\em Phys. Rev. C}, 97(4):041902, 2018.

\bibitem{Xie:2017upb}
Yilong Xie, Dujuan Wang, and L\'aszl\'o~P. Csernai.
\newblock {Global \ensuremath{\Lambda} polarization in high energy collisions}.
\newblock {\em Phys. Rev. C}, 95(3):031901, 2017.

\bibitem{Florkowski:2017ruc}
Wojciech Florkowski, Bengt Friman, Amaresh Jaiswal, and Enrico Speranza.
\newblock {Relativistic fluid dynamics with spin}.
\newblock {\em Phys. Rev. C}, 97(4):041901, 2018.

\bibitem{Hattori:2019lfp}
Koichi Hattori, Masaru Hongo, Xu-Guang Huang, Mamoru Matsuo, and Hidetoshi
  Taya.
\newblock {Fate of spin polarization in a relativistic fluid: An
  entropy-current analysis}.
\newblock {\em Phys. Lett. B}, 795:100--106, 2019.

\bibitem{Xie:2019wxz}
Yilong Xie, Gang Chen, and Laszlo~Pal Csernai.
\newblock {A study of $\Lambda $ and $\bar{\Lambda }$ polarization splitting by
  meson field in PICR hydrodynamic model}.
\newblock {\em Eur. Phys. J. C}, 81(1):12, 2021.

\bibitem{Fukushima:2020ucl}
Kenji Fukushima and Shi Pu.
\newblock {Spin hydrodynamics and symmetric energy-momentum tensors
  \textendash{} A current induced by the spin vorticity \textendash{}}.
\newblock {\em Phys. Lett. B}, 817:136346, 2021.

\bibitem{Li:2020eon}
Shiyong Li, Mikhail~A. Stephanov, and Ho-Ung Yee.
\newblock {Nondissipative Second-Order Transport, Spin, and Pseudogauge
  Transformations in Hydrodynamics}.
\newblock {\em Phys. Rev. Lett.}, 127(8):082302, 2021.

\bibitem{Fu:2020oxj}
Baochi Fu, Kai Xu, Xu-Guang Huang, and Huichao Song.
\newblock {Hydrodynamic study of hyperon spin polarization in relativistic
  heavy ion collisions}.
\newblock {\em Phys. Rev. C}, 103(2):024903, 2021.

\bibitem{Bhadury:2021oat}
Samapan Bhadury, Jitesh Bhatt, Amaresh Jaiswal, and Avdhesh Kumar.
\newblock {New developments in relativistic fluid dynamics with spin}.
\newblock {\em Eur. Phys. J. ST}, 230(3):655--672, 2021.

\bibitem{Ryu:2021lnx}
Sangwook Ryu, Vahidin Jupic, and Chun Shen.
\newblock {Probing early-time longitudinal dynamics with the
  \ensuremath{\Lambda} hyperon's spin polarization in relativistic heavy-ion
  collisions}.
\newblock {\em Phys. Rev. C}, 104(5):054908, 2021.

\bibitem{Wu:2022mkr}
Xiang-Yu Wu, Cong Yi, Guang-You Qin, and Shi Pu.
\newblock {Local and global polarization of \ensuremath{\Lambda} hyperons
  across RHIC-BES energies: The roles of spin hall effect, initial condition,
  and baryon diffusion}.
\newblock {\em Phys. Rev. C}, 105(6):064909, 2022.

\bibitem{Sun:2017xhx}
Yifeng Sun and Che~Ming Ko.
\newblock {$\Lambda$ hyperon polarization in relativistic heavy ion collisions
  from a chiral kinetic approach}.
\newblock {\em Phys. Rev. C}, 96(2):024906, 2017.

\bibitem{Liu:2019krs}
Shuai Y.~F. Liu, Yifeng Sun, and Che~Ming Ko.
\newblock {Spin Polarizations in a Covariant Angular-Momentum-Conserved Chiral
  Transport Model}.
\newblock {\em Phys. Rev. Lett.}, 125(6):062301, 2020.

\bibitem{Wang:2020pej}
Ziyue Wang, Xingyu Guo, and Pengfei Zhuang.
\newblock {Equilibrium Spin Distribution From Detailed Balance}.
\newblock {\em Eur. Phys. J. C}, 81(9):799, 2021.

\bibitem{Yang:2020hri}
Di-Lun Yang, Koichi Hattori, and Yoshimasa Hidaka.
\newblock {Effective quantum kinetic theory for spin transport of fermions with
  collsional effects}.
\newblock {\em JHEP}, 07:070, 2020.

\bibitem{Weickgenannt:2020aaf}
Nora Weickgenannt, Enrico Speranza, Xin-li Sheng, Qun Wang, and Dirk~H.
  Rischke.
\newblock {Generating Spin Polarization from Vorticity through Nonlocal
  Collisions}.
\newblock {\em Phys. Rev. Lett.}, 127(5):052301, 2021.

\bibitem{Huang:2011ru}
Xu-Guang Huang, Pasi Huovinen, and Xin-Nian Wang.
\newblock {Quark Polarization in a Viscous Quark-Gluon Plasma}.
\newblock {\em Phys. Rev. C}, 84:054910, 2011.

\bibitem{Pang:2018zzo}
Long-Gang Pang, H.~Petersen, and Xin-Nian Wang.
\newblock {Pseudorapidity distribution and decorrelation of anisotropic flow
  within the open-computing-language implementation CLVisc hydrodynamics}.
\newblock {\em Phys. Rev. C}, 97(6):064918, 2018.

\bibitem{Wu:2018cpc}
Xiang-Yu Wu, Long-Gang Pang, Guang-You Qin, and Xin-Nian Wang.
\newblock {Longitudinal fluctuations and decorrelations of anisotropic flows at
  energies available at the CERN Large Hadron Collider and at the BNL
  Relativistic Heavy Ion Collider}.
\newblock {\em Phys. Rev. C}, 98(2):024913, 2018.

\bibitem{STAR:2019erd}
Jaroslav Adam et~al.
\newblock {Polarization of $\Lambda$ ($\bar{\Lambda}$) hyperons along the beam
  direction in Au+Au collisions at $\sqrt{s_{_{NN}}}$ = 200 GeV}.
\newblock {\em Phys. Rev. Lett.}, 123(13):132301, 2019.

\bibitem{Fu:2021pok}
Baochi Fu, Shuai Y.~F. Liu, Longgang Pang, Huichao Song, and Yi~Yin.
\newblock {Shear-Induced Spin Polarization in Heavy-Ion Collisions}.
\newblock {\em Phys. Rev. Lett.}, 127(14):142301, 2021.

\bibitem{Becattini:2021iol}
F.~Becattini, M.~Buzzegoli, G.~Inghirami, I.~Karpenko, and A.~Palermo.
\newblock {Local Polarization and Isothermal Local Equilibrium in Relativistic
  Heavy Ion Collisions}.
\newblock {\em Phys. Rev. Lett.}, 127(27):272302, 2021.

\bibitem{Loizides:2017ack}
C.~Loizides, J.~Kamin, and D.~d'Enterria.
\newblock {Improved Monte Carlo Glauber predictions at present and future
  nuclear colliders}.
\newblock {\em Phys. Rev. C}, 97(5):054910, 2018.
\newblock [Erratum: Phys.Rev.C 99, 019901 (2019)].

\bibitem{Karpenko:2021wdm}
Iurii Karpenko.
\newblock {Vorticity and Polarization in Heavy Ion Collisions: Hydrodynamic
  Models}.
\newblock {\em arXiv:2101.04963}, 2021.

\bibitem{Danielewicz:1984ww}
P.~Danielewicz and M.~Gyulassy.
\newblock {Dissipative Phenomena in Quark Gluon Plasmas}.
\newblock {\em Phys. Rev. D}, 31:53--62, 1985.

\bibitem{Jiang:2021ajc}
Ze-Fang Jiang, Shanshan Cao, Xiang-Yu Wu, C.~B. Yang, and Ben-Wei Zhang.
\newblock {Longitudinal distribution of initial energy density and directed
  flow of charged particles in relativistic heavy-ion collisions}.
\newblock {\em Phys. Rev. C}, 105(3):034901, 2022.

\bibitem{Bozek:2010bi}
P.~Bozek and I.~Wyskiel.
\newblock {Directed flow in ultrarelativistic heavy-ion collisions}.
\newblock {\em Phys. Rev. C}, 81:054902, 2010.

\bibitem{Bozek:2011ua}
Piotr Bozek.
\newblock {Flow and interferometry in 3+1 dimensional viscous hydrodynamics}.
\newblock {\em Phys. Rev. C}, 85:034901, 2012.

\bibitem{Jiang:2021foj}
Ze-Fang Jiang, C.~B. Yang, and Qi~Peng.
\newblock {Directed flow of charged particles within idealized viscous
  hydrodynamics at energies available at the BNL Relativistic Heavy Ion
  Collider and at the CERN Large Hadron Collider}.
\newblock {\em Phys. Rev. C}, 104(6):064903, 2021.

\bibitem{Shen:2020jwv}
Chun Shen and S.~Alzhrani.
\newblock {Collision-geometry-based 3D initial condition for relativistic
  heavy-ion collisions}.
\newblock {\em Phys. Rev. C}, 102(1):014909, 2020.

\bibitem{Borsanyi:2013bia}
S.~Borsanyi, Z.~Fodor, C.~Hoelbling, S.~D. Katz, S.~Krieg, and K.~K. Szabo.
\newblock {Full result for the QCD equation of state with 2+1 flavors}.
\newblock {\em Phys. Lett. B}, 730:99--104, 2014.

\bibitem{Becattini:2015ska}
F.~Becattini, G.~Inghirami, V.~Rolando, A.~Beraudo, L.~Del~Zanna, A.~De~Pace,
  M.~Nardi, G.~Pagliara, and V.~Chandra.
\newblock {A study of vorticity formation in high energy nuclear collisions}.
\newblock {\em Eur. Phys. J. C}, 75(9):406, 2015.
\newblock [Erratum: Eur.Phys.J.C 78, 354 (2018)].

\bibitem{Sheng:2020ghv}
Xin-Li Sheng, Qun Wang, and Xin-Nian Wang.
\newblock {Improved quark coalescence model for spin alignment and polarization
  of hadrons}.
\newblock {\em Phys. Rev. D}, 102(5):056013, 2020.

\end{thebibliography}
	
\end{document}